\pdfoutput=1
\newcommand*{\ATLASLATEXPATH}{./}

\documentclass[american,texlive=2016,cernpreprint]{\ATLASLATEXPATH atlasdoc}

\usepackage[subcaption,backend=biber]{atlaspackage}
\usepackage{atlasbiblatex}

\usepackage{atlasphysics}

\usepackage{EXOT-2017-05-PAPER-defs}

\usepackage{multirow}
\usepackage{bigstrut}
\usepackage{pdflscape}
\usepackage{changepage}

\AtlasTitle{Search for long-lived particles produced in $pp$ collisions at $\sqrt{s}=13$~\TeV\ that decay into displaced hadronic jets in the ATLAS muon spectrometer }

\AtlasAbstract{%
A search for the decay of neutral, weakly interacting, long-lived particles using data collected by the ATLAS detector at the LHC is presented. The analysis in this paper uses 36.1~fb$^{-1}$ of proton--proton collision data at $\sqrt{s} = 13$~\TeV\ recorded in 2015--2016. The search employs techniques for reconstructing vertices of long-lived particles decaying into jets in the muon spectrometer exploiting a two vertex strategy and a novel technique that requires only one vertex in association with additional activity in the detector that improves the sensitivity for longer lifetimes. The observed numbers of events are consistent with the expected background and limits for several benchmark signals are determined.
}

\PreprintIdNumber{CERN-EP-2018-241}
\AtlasJournal{Phys.\ Rev.\ D}

\hypersetup{pdftitle={ATLAS document},pdfauthor={The ATLAS Collaboration}}

\begin{document}

\maketitle

\tableofcontents	

\clearpage

\author{The ATLAS Collaboration}
\section{Introduction}\label{sec:intro}
The discovery of the Higgs boson at the LHC completed the Standard Model (SM) of elementary particles and focused attention on the many central features of our universe that the SM does not address: dark matter (DM), neutrino mass, matter--antimatter asymmetry (baryogenesis), and the hierarchy problem (naturalness). Many Beyond the Standard Model (BSM) theoretical constructs proposed in the past few years that address these phenomena predict the existence of long-lived particles (LLPs) with macroscopic decay lengths that are limited only by Big Bang Nucleosynthesis to about $c \tau \lesssim 10^7$--$10^8$~m, where $\tau$ is the proper lifetime of the LLP \cite{BBN}. Examples include: supersymmetric (SUSY) models such as mini split SUSY \cite{Arvanitaki:2012ps, ArkaniHamed:2012gw}, gauge-mediated SUSY breaking \cite{Giudice:1998bp}, $R$-parity-violating (RPV) SUSY \cite{Barbier:2004ez, Csaki:2013jza} and Stealth SUSY \cite{Josh1,Josh2}; models addressing the hierarchy problem such as Neutral Naturalness \cite{Chacko:2015fbc,Burdman:2006tz, Cai:2008au, Chacko:2005pe} and Hidden Valleys~\cite{Strassler:2006im,Strassler:2006ri}; models addressing dark matter \cite{Baumgart:2009tn, Kaplan:2009ag, Chan:2011aa, Dienes:2011ja, Dienes:2012yz}, and the \mbox{matter--antimatter} asymmetry of the universe \cite{Cui:2014twa}; and models that generate neutrino masses \cite{Helo:2013esa,Batell:2016zod}.  Many of these theoretical models result in neutral LLPs, which may be produced in the proton--proton collisions of the LHC and decay back into SM particles far from the interaction point (IP).

Searches for LLPs decaying into final states containing jets were carried out at the Tevatron ($\sqrt{s}=1.96$~\TeV) by both the CDF~\cite{CDF} and D0~\cite{D0} Collaborations, at the LHC by the ATLAS and LHCb Collaborations in proton--proton collisions at $\sqrt{s}=7$~\TeV~\cite{HVAtlas,LHCb}, by the ATLAS, CMS and LHCb Collaborations at $\sqrt{s}=8$~\TeV~\cite{HVAtlas8TeV,SUSY-2014-02,CMS,CalRatio, LHCb-PAPER-2016-065,LHCb-1612.00945} and more recently by the CMS Collaboration at $\sqrt{s}=13$~\TeV~\cite{CMS-EXO-16-003}. 
To date, no search has observed evidence of BSM, neutral LLPs.

This paper describes a search for neutral LLPs produced in proton--proton interactions at $\sqrt{s} = 13$~\TeV, using 36.1~fb$^{-1}$ of data recorded by the ATLAS detector at the LHC during 2015 and 2016.  Decays of LLPs can result in secondary decay vertices (displaced vertices) that are highly displaced from the IP. The present paper focuses on LLP decays reconstructed in the muon spectrometer. Three different analysis strategies are considered, with each strategy targeting a specific event topology. Two single-vertex strategies are based on methodologies presented in Ref.~\cite{LLPDataDriven}, and the third strategy is an inclusive search for two displaced vertices in the muon spectrometer.

This work significantly extends the mean proper lifetime ($c\tau$) range of the ATLAS search for a light scalar boson decaying into long-lived neutral particles beyond that at $\sqrt{s}=8$~\TeV\ in 20.3~fb$^{-1}$ of 2012 proton--proton collision data~\cite{HVAtlas8TeV}, which covered the $c\tau$ region 1 to 100~m.  Additionally, it extends the range of excluded proper lifetimes beyond that of a recent ATLAS analysis~\cite{CalRatio} that searches for displaced decays in the hadronic calorimeter and uses the same scalar boson model and mass points.

The paper first describes the ATLAS detector in Section~\ref{sec:ATLASdetector}, followed by the event selection strategy in Section~\ref{sec:strategy}, the benchmark models in Section~\ref{sec:models}, and the data and simulation samples in Section~\ref{sec:samples}. The specialized trigger and reconstruction algorithms are discussed in Section~\ref{sec:trigreco}, followed by a description of the baseline selection applied to all events in Section~\ref{sec:baseselection}. The next two sections (\ref{sec:2msvx} and \ref{sec:1msvx}) outline the three search topologies. Systematic uncertainties are summarized in Section~\ref{sec:systematics} and results for all three topologies are presented in Section~\ref{sec:result}. The summary and conclusions are given in Section~\ref{sec:conclusion}.

\section{ATLAS detector}\label{sec:ATLASdetector}
The ATLAS detector~\cite{atlas-detector}, which has nearly 4$\pi$ steradian coverage, is a multipurpose detector consisting of an inner tracking detector (ID) surrounded by a superconducting solenoid, electromagnetic and hadronic calorimeters, and a muon spectrometer (MS) based on three large air-core toroidal superconducting magnets, each with eight coils. The ID covers the range 0.03~m~$<r<$~1.1~m and $|z|<3.5$~m.\footnote{ATLAS uses a right-handed coordinate system with its origin at the nominal IP in the center of the detector and the $z$-axis along the beam pipe. The $x$-axis points from the IP to the center of the LHC ring, and the $y$-axis points upwards. Cylindrical coordinates $(r,\phi)$ are used in the transverse plane, where $\phi$ is the azimuthal angle around the $z$-axis. The pseudorapidity is defined in terms of the polar angle $\theta$ as $\eta = - \ln \tan(\theta/2)$. Angular distance is measured in units of $\Delta R \equiv \sqrt{(\Delta\eta)^{2} + (\Delta\phi)^{2}}$.} It consists of a silicon pixel detector, a silicon microstrip detector, and a straw-tube transition-radiation tracker.  Together, the three systems provide precision tracking of charged particles for \mbox{$|\eta|<2.5$}.

The calorimeter system covers the pseudorapidity range $|\eta| < 4.9$. It consists of a high-granularity lead/liquid-argon electromagnetic calorimeter (ECal) surrounded by a hadronic calorimeter (HCal).
Within the region $|\eta|< 3.2$, the \ECal comprises liquid-argon (LAr) barrel and endcap electromagnetic calorimeters with lead absorbers. An additional thin LAr presampler covering $|\eta| < 1.8$ is used to correct for energy loss in material upstream of the calorimeters. The \ECal extends from 1.5~m to 2.0~m in $r$ in the barrel and from 3.6~m to 4.25~m in $|z|$ in the \Endcaps. 
The \HCal is a steel/scintillator-tile calorimeter that is segmented into three barrel structures within $|\eta| < 1.7$, and two copper/LAr hadronic calorimeters in the endcap ($1.5 < |\eta| < 3.2$). The \HCal covers the region from 2.25~m to 4.25~m in $r$ in the barrel (although the HCal active material extends only up to 3.9 m) and from 4.3~m to 6.05~m in $|z|$ in the \Endcaps. The solid angle coverage is completed with forward copper/LAr and tungsten/LAr calorimeter modules optimized for electromagnetic and hadronic measurements, respectively. Together the ECal and HCal have a thickness of 9.7 interaction lengths at $\eta=0$.

The MS comprises three stations of separate trigger and tracking chambers that measure the deflection of muons in a magnetic field generated by the air-core toroid magnets. The barrel chamber system is subdivided into 16 sectors: 8 large sectors (between the magnet coils) and 8 small sectors (inside the magnet coils).
Three stations of resistive plate chambers (RPC) and thin gap chambers (TGC) are used for triggering in the MS barrel and \Endcaps, respectively. The first two RPC stations, which are radially separated by 0.5~m, start at a radius of either 7~m (large sectors) or 8~m (small sectors). The third station is located at a radius of either 9~m (large sectors) or 10~m (small sectors). In the \Endcaps, the first TGC station is located at $|z|=13$~m. The other two stations start at $|z|=14$~m and $|z|=14.5$~m, respectively. The muon trigger system covers the range $|\eta| < 2.4$. 
The muon tracking chamber system covers the region $|\eta| < 2.7$ with three layers of monitored drift tubes (MDT), complemented by cathode strip chambers (CSC) in the forward region. The MDT chambers consist of two multilayers separated by a distance ranging from 6.5~mm to 317~mm. Each multilayer consists of three or four layers of drift tubes. The individual drift tubes are 30~mm in diameter and have a length of 2--5~m depending on the location of the chamber in the spectrometer. 
In each multilayer the charged-particle track segment can be reconstructed by finding the line that is tangent to the drift circles. These segments are local measurements of the position and direction of the charged particle. Because the tubes are 2--5~m in length with a direction along $\phi$, the MDT measurement provides only a very coarse $\phi$ position of the track hit. In order to reconstruct the $\phi$ position and direction, the MDT measurements are combined with the $\phi$ coordinate measurements from the trigger chambers.

The ATLAS trigger and data acquisition system~\cite{TRIG-2016-01} consists of a hardware-based first-level trigger (L1) followed by a software-based high-level trigger (HLT) that reduces the rate of recorded events for offline storage to 1~kHz.

The implementation of the L1 muon trigger logic is similar for both the RPC and TGC systems. Each of the three planes of the RPC system and the two outermost planes of the TGC system consist of a doublet of independent detector layers.  The first TGC plane contains three detector layers. A low-$\pT$ ($< 10$~\GeV) muon region-of-interest (\roi) is generated by requiring a coincidence of hits in at least three of the four layers of the two inner RPC planes for the barrel. In the endcaps, the trigger requires hits in the two outer TGC planes. A high-$\pT$ muon \roi requires additional hits in at least one of the two layers of the outer RPC plane for the barrel, while for the endcaps, hits in two of the three layers of the innermost TGC layer are required.
The muon \rois have a spatial extent of $0.2\times0.2$ in $\Delta\eta \times \Delta\phi$ in the MS barrel and $0.1\times0.1$ in $\Delta\eta \times \Delta\phi$ in the MS \Endcaps. Only the two highest-$\pT$ \rois per MS sector are used by the HLT.

The L1 calorimeter trigger is based on information from the calorimeter elements within projective regions, called trigger towers. The trigger towers have a size of approximately 0.1 in $\Delta\eta$ and $\Delta\phi$ in the central part of the calorimeter, \mbox{$|\eta|<2.5$}, and are larger and less uniform in the more forward region.

\section{Analysis strategy}\label{sec:strategy}

The analysis presented in this paper searches for events with two displaced vertices in the MS, or one displaced vertex in the MS in association with additional activity in the detector. Three separate strategies are studied, defined by the number of MS vertices and additional selection criteria. The benchmark models that motivate these strategies are discussed in detail in Section~\ref{sec:models}.

Candidate events are selected by the Muon RoI Cluster trigger~\cite{MuonRoITrigger} that requires a cluster of three (four) muon RoIs in the barrel (endcaps). No jet or track isolation requirements are applied at trigger level. Displaced vertices are then reconstructed using a dedicated MS vertex reconstruction algorithm~\cite{MSVXRECO}. 

The simplest strategy requires at least two MS vertices (2MSVx), and it is inclusive of any other activity in the event. The other two strategies require exactly one MS vertex, with additional requirements on associated objects (1MSVx+AO). The first requires exactly one MS vertex and two prompt jets (1MSVx+Jets), targeting models where prompt jets are produced together with the LLP, as expected in the Stealth SUSY scenarios. In these cases, the two prompt jets can also contribute to signal event selection. The second requires a small amount of missing transverse momentum (denoted by \MET) in addition to the single displaced vertex (1MSVx+\MET) and targets models that do not predict significant activity in addition to LLPs, such as from decays of ``SM-like'' Higgs bosons into long-lived neutral scalar particle pairs. In fact, since the tracks originating from vertices in the MS are not considered in the \MET computation, for these models the \MET in signal events is sensitive to the Higgs boson \pT, which is typically of the order of tens of GeV. For signal vertices, if the second LLP has decayed in the ID or calorimeter the \MET vector tends to be aligned with the direction of the displaced vertex, measured from the origin of the detector coordinate system. However, if the decay of the second LLP occur beyond the MS there is no missing energy. Therefore, the angle in the transverse plane between the vertex and the \MET direction can contribute to the signal event selection. The three analysis strategies are summarized in Table~\ref{tab:topo},  together with the theoretical benchmark models used in this paper.
 
\begin{table}[hbtp]
\centering
\caption{Topologies considered in this paper, corresponding basic event selection and benchmark models.}
\begin{tabular}{lll}
\toprule
\multicolumn{1}{c}{Strategy} &\multicolumn{1}{c}{Basic event selection} & \multicolumn{1}{c}{Benchmarks} \\
\midrule
\midrule
\multirow{2}{*}{2MSVx} & \multirow{2}{*}{At least 2 MS vertices} &   Scalar portal, Higgs portal baryogenesis,\\
                                      &                                                             &   Stealth SUSY \\
\midrule
\multirow{2}{*}{1MSVx+Jets} & Exactly 1 MS vertex & \multirow{2}{*}{Stealth SUSY} \\
					    & At least 2 jets with $\ET > 150$~\GeV\ & \\
 \midrule
\multirow{2}{*}{1MSVx+\MET} &  Exactly 1 MS vertex& Scalar portal with $m_{\varPhi}=125$~\GeV, \\
				  	      &  $\MET > 30$~\GeV\ & Higgs portal baryogenesis \\
\bottomrule
\end{tabular}
\label{tab:topo}
\end{table}

The main source of background to LLPs decaying into hadronic jets in the MS is from hadronic or electromagnetic showers not contained in the calorimeter volume (\emph{punch-through jets}) resulting in tracks reconstructed in the MS. Multijet events that contain vertices in the MS would have ID tracks and jets that point towards the displaced MS vertex as well as inwards to the IP.  To reduce the acceptance of fake vertices from multijet events, vertices are required to be isolated from ID tracks and calorimeter jets. Additional background, referred to in this paper as \emph{non-collision background}, can be generated by electronic noise in the MDT and RPC/TGC chambers, by cosmic-ray muons, by multijet events with mismeasured jets and by machine-induced background~\cite{BIB}. This last contribution, usually referred to as \emph{beam-induced background}, is composed of particles produced in the hadronic and electromagnetic showers caused by beam protons interacting with collimators or residual gas molecules inside the vacuum pipe.

To avoid unintended biasing of the results, the signal regions of the 2MSVx and 1MSVx+AO strategies were blinded during the analysis development.

\section{Description of benchmark models}\label{sec:models}

\begin{figure}[thbp]
\centering
\begin{subfigure}[b]{0.35\textwidth}
\includegraphics[width=\textwidth]{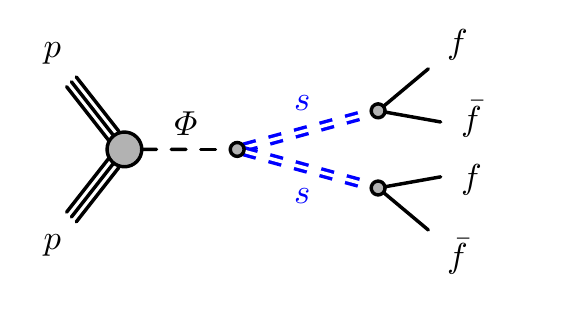}
\caption{}\label{fig:HiggsFeyn}
\end{subfigure}	
\begin{subfigure}[b]{0.32\textwidth} \hspace{-0.5 cm}
\includegraphics[width=\textwidth]{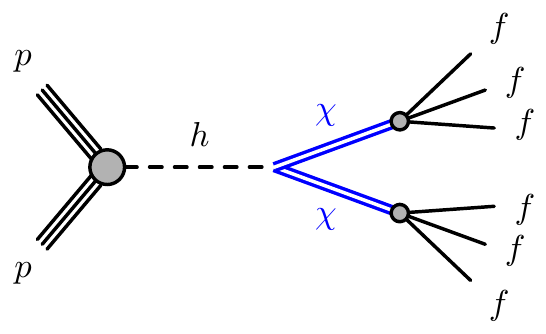}
\caption{}\label{fig:BaryogenesisFeyn}
\end{subfigure}
\begin{subfigure}[b]{0.3\textwidth}
\includegraphics[width=\textwidth]{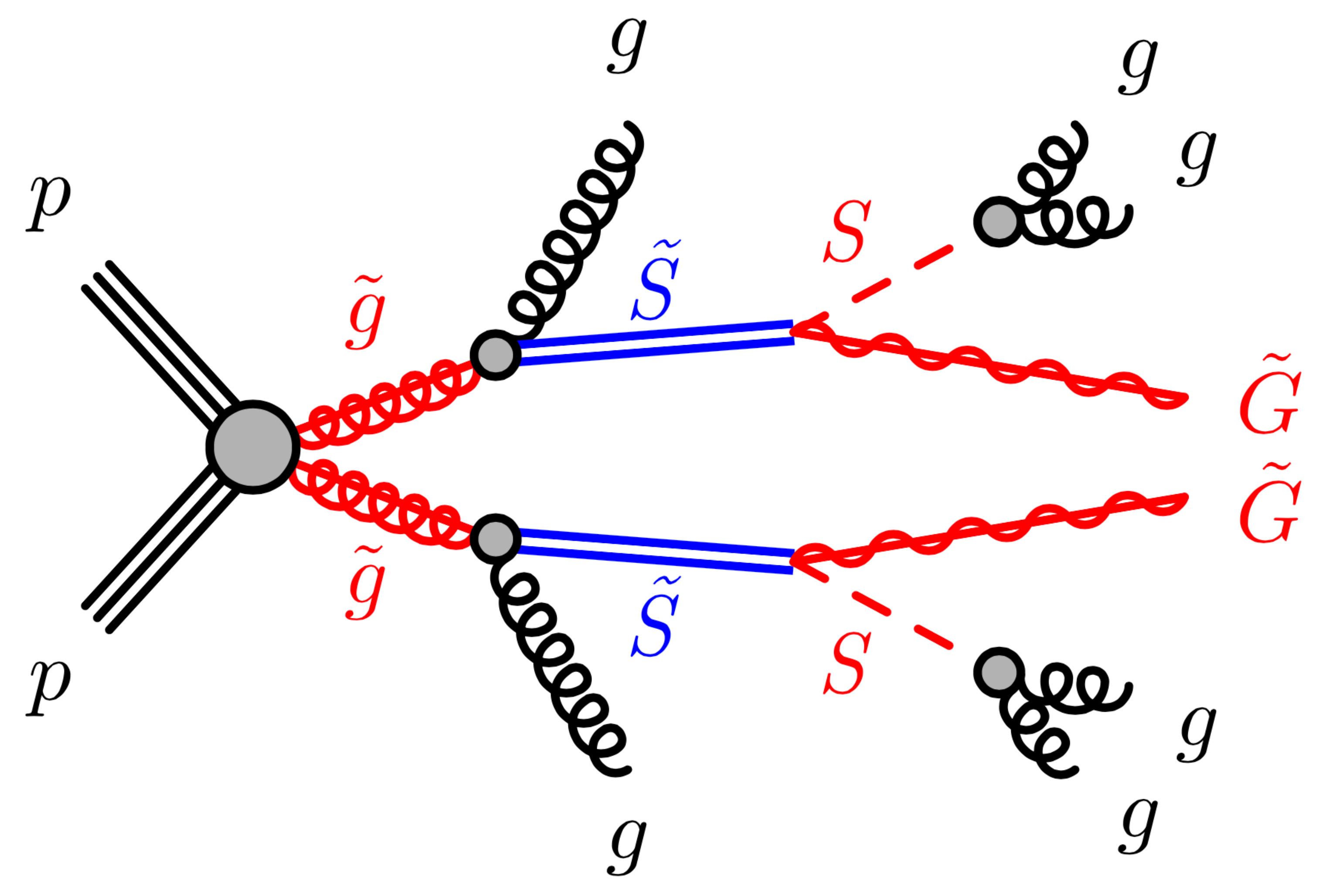}
\caption{}\label{fig:StealthFeyn}
\end{subfigure}
\caption{
Diagrams of the benchmark models studied in this paper: (a) scalar portal model, (b) Higgs portal baryogenesis model, and (c) Stealth SUSY model. The LLPs in these processes are represented by double lines and labeled (a) $s$, (b) $\chi$, and (c) \textit{\~{S}}. In the Stealth SUSY model, \textit{\~{G}} is the gravitino and $S$ is the singlet. The final-state SM fermions are labeled as $f$, and the gluons as $g$.}
\label{fig:Diagrams}
\end{figure}

Although the event selections outlined in Section \ref{sec:strategy} are sensitive to a large variety of models, this paper interprets the results in terms of three different benchmark models. The first, shown in Figure~\ref{fig:HiggsFeyn}, is a \emph{scalar portal} model~\cite{Strassler:2006ri}, where a SM-like Higgs or lower/higher-mass boson ($\varPhi$) decays into two long-lived scalars ($s$). Figure~\ref{fig:BaryogenesisFeyn} shows the second model, \emph{Higgs portal baryogenesis}~\cite{Cui:2014twa}, in which a SM-like Higgs boson ($h$) decays into long-lived Majorana fermions $\chi$ that decay into fermions, violating baryon and/or lepton number conservation. The last model, shown in Figure~\ref{fig:StealthFeyn}, is a \emph{Stealth SUSY} model~\cite{Josh1,Josh2} where the long-lived singlino ($\tilde{S}$) is produced by a gluino ($\tilde{g}$) in association with a prompt gluon-jet ($g$). The singlino decay produces two gluons and a light gravitino.

The decay channels, the relative masses and lifetimes generated for each model, as well as details about the Monte Carlo (MC) event generation are described in Section~\ref{sec:samples}.

\subsection{Scalar portal}\label{sec:HSSmodel}

A theoretically popular way of introducing long-lived, neutral particles to the SM is through a hidden sector that weakly couples to the SM. For example, scalar (Higgs) portals\,\cite{Strassler:2006im, Strassler:2006ri,Strassler:2006qa}, where the Higgs boson weakly mixes with a hidden-sector scalar, can result in pair production of hidden-sector scalars or pseudoscalars that carry no SM quantum numbers. The branching fraction limit for SM Higgs boson decays into undetected particles is
currently at the 25\% level~\cite{HIGG-2015-07} (assuming SM-like Higgs boson production and width), potentially allowing sizable branching fractions for decays into non-SM particles.

Moreover, models of neutral naturalness~\cite{Uncolored} are generic extensions of hidden-valley portal models where the scalar masses can be very low, typically 5 to 15~\GeV. To date, no LHC analysis has explored this model. 

The mechanism for LLP production in scalar decays is shown in Figure~\ref{fig:HiggsFeyn}. Here, a scalar boson $\varPhi$ decays with some effective coupling into a pair of long-lived scalars, $s$. The scalars $s$ subsequently decay into SM particles. 
Since this model assumes that the couplings of the scalar to SM particles are determined by a Yukawa coupling, each long-lived scalar decays mainly into heavy fermions, $b\overline{b}$, $c\overline{c}$, and $\tau^{+}\tau^{-}$. The branching fractions of these decays depend on the mass of the scalar, $m_s$, but for $m_s \gtrsim 25$~\GeV\ they are almost constant and equal to 85\%, 5\%, and 8\%.

The branching fraction for $\varPhi$ decaying into a pair of hidden-sector particles is not constrained in these models. It is therefore interesting to focus both on Higgs boson decays into LLPs, where $\varPhi$ is a SM-like Higgs boson, and on other $\varPhi$ mass regions previously unexplored for decays into LLPs.

\subsection{Higgs portal baryogenesis}\label{sec:Baryogenesismodel}

The origin of the cosmic asymmetric abundance of baryons remains one of the most prominent questions that demand physics beyond the SM. Several baryogenesis mechanisms have been proposed, but electroweak baryogenesis is one of the few with signatures that could be explored at the LHC energies. In addition to better testability, baryogenesis based on new weak-scale particles is also theoretically appealing since it can naturally connect new physics addressing the weak-scale hierarchy problem with the dynamics responsible for generating the baryon asymmetry. A few examples of low-scale ($\lesssim$~\TeV) baryogenesis models that generate the baryon asymmetry via the decays of weak-scale states have been shown to have direct testability at colliders~\cite{Leptogenesis1,Leptogenesis2,Baryogenesis1}.

In the baryogenesis model considered for this paper~\cite{Cui:2014twa}, the lowest-dimension operator coupling a singlet $\chi$ to the SM is the Higgs portal. The simplest realization of this interaction is with a scalar, $\varPhi$, that mixes with the SM Higgs boson~\cite{NonSMCoupling2}. If $\varPhi$ has a Yukawa coupling to a pair of $\chi$, this leads to the Higgs portal production of $\chi$ via exchange of a single SM-like Higgs boson after mixing, $pp \rightarrow h \rightarrow \chi\chi$, as shown in Figure~\ref{fig:BaryogenesisFeyn}. 
Since LHC experiments have established the existence of a SM-like Higgs boson with a mass of 125~\GeV, while the other possibilities are more model-dependent, the model used here assumes the minimal spectrum where the $\varPhi$ scalar is heavy and decouples, and focuses on the production channel via the SM Higgs portal.

For the production of the $\chi$ through the Higgs portal, two different regimes can be identified.

\begin{itemize}
\item $m_\chi < m_h/2$: in this region the dominant production mechanism is through an \emph{on-shell} Higgs boson. The $\chi$ production at 13~\TeV\ is expected to be copious, 
$O(10~\textrm{pb})$, and the constraints set by the current LHC searches are correspondingly strong. There are also indirect limits on the non-SM decay branching fraction of the Higgs boson based on global fits~\cite{nonSMHiggsConstr1,nonSMHiggsConstr2}. Despite all these strong constraints, the on-shell region is still very interesting due to the sizable branching fractions allowed for the Higgs decays into BSM particles~\cite{nonSMHiggsConstr1,nonSMHiggsConstr2}.
\item $m_\chi > m_h/2$: in this region the Higgs boson is \emph{off-shell} and the signal rate falls rapidly with increasing $m_\chi$, even for large mixing. The cross section expected for a $\chi$ mass of 100~\GeV\ is about 7~fb. 
\end{itemize}

The decay modes of the $\chi$ must violate baryon and/or lepton number conservation, which generates the baryonic asymmetry. The lowest-dimensional interactions of this type allow $\chi$ to decay into three SM fermions. The decay channels used in this paper, $\chi \rightarrow \tau^+ \, \tau^- \, \nu_\ell$ , $c \, b \, s$ , $\ell^+ \, \bar{c} \, b$ ,  $\nu \, b \, \bar{b}$, are examples of three types of couplings inspired by $R$-parity-violating SM fermion trilinear operators that can couple to $\chi$~\cite{Cui:2014twa}. The charge conjugates of these decay channels are also considered. Decays into final states as $c \, b \, b$ or $c \, s \, s$ are not possible since the $R$-parity-violating operator containing two down-type quarks with the same flavor is not allowed. Mixed cases such as $c \, b \, d$ or $c \, d \, s$ are possible, but they are not considered in this paper since their kinematics is similar to the $c \, b \, s$ channel and give similar results.

\subsection{Stealth SUSY}\label{sec:StealthModel}

Stealth SUSY models~\cite{Josh1,Josh2} are a class of $R$-parity-conserving SUSY models that do not have large \MET signatures. While this can be accomplished in many different ways, this search explores a model that involves adding a hidden-sector (stealth) singlet superfield $S$ at the electroweak scale, which has a superpartner singlino $\tilde{S}$. By weakly coupling the hidden sector to the Minimal Supersymmetric Standard Model (MSSM)~\cite{MSSMmodel}, the mass-splitting between $S$ and $\tilde{S}$ ($\delta M$) is small, assuming low-scale SUSY breaking. High-scale SUSY breaking also can be consistent with small mass splitting and Stealth SUSY, although this requires a more complex model and is not considered in this search~\cite{Josh2}. 

The SUSY decay chain ends with the singlino decaying into a singlet plus a low-mass gravitino $\tilde{G}$, where the gravitino carries off very little energy and the singlet promptly decays into two gluons. The effective decay processes are $\tilde{g} \rightarrow \tilde{S} g$ (prompt), $\tilde{S} \rightarrow S \tilde{G}$ (not prompt), and $S \rightarrow gg$ (prompt), where the gravitino is treated as massless. This scenario results in one prompt gluon and two displaced gluons per gluino decay. Since $R$-parity is assumed to be conserved, each event necessarily produces two gluinos, resulting in two displaced vertices. A representative diagram of this process is shown in Figure~\ref{fig:StealthFeyn}. The simplified Stealth SUSY model considered in this paper assumes that all squarks are decoupled.

The decay width (and, consequently, the lifetime) of the singlino is determined by both the $\delta M$ and the SUSY-breaking scale $\sqrt{F}$: ${\Gamma}_{\tilde{S}~\rightarrow~S \tilde{G}}~\approx~m_{\tilde{S}}(\delta M)^4/\pi F^2$ ~\cite{Josh1}. The SUSY-breaking scale $\sqrt{F}$ is not a fixed parameter, and thus the singlino has the possibility of traveling an appreciable distance through the detector, leading to a significantly displaced vertex. 

\section{Data and simulation samples}\label{sec:samples}

The analysis presented in this paper uses $\sqrt{s} = 13$~\TeV\ $pp$ collision data recorded by the ATLAS detector with stable LHC beams during the 2015 and 2016 data-taking periods. After data quality requirements, the total integrated luminosity is 3.2~fb$^{-1}$ and 32.9~fb$^{-1}$ for 2015 and 2016, respectively. 

\begin{table}[h!]
\centering
\caption{Mass parameters for the simulated scalar portal, Higgs portal baryogenesis, and Stealth SUSY models.}
\begin{tabular}{ccl}
\toprule
Model & $m_\phi$ [\GeV] & $m_s$ [\GeV] \\
\hline
\multirow{6}{*}{Scalar portal} & 100 & 8 , 25 \\
                          & 125 & 5 , 8 , 15 , 25 , 40 \\
                          & 200 & 8 , 25 , 50 \\
                          & 400 & 50 , 100 \\
                          & 600 & 50 , 150 \\
                          & 1000 & 50 , 150 , 400 \\
\hline\hline
& $m_\chi$ [\GeV] & $\chi$ decay channel \\
\hline
\multirow{4}{*}{Higgs portal baryogenesis} & 10 & \multirow{4}{*}{$\tau^+ \, \tau^- \, \nu_\ell$ , $c \, b \, s$ , $\ell^+ \, \bar{c} \, b$ ,  $\nu \, b \, \bar{b}$} \\
                          & 30 & \\
                          & 50 & \\
                          & 100 & \\
\hline\hline
& $m_{\tilde{g}}$ [\GeV] & $m_{\tilde{S}}$, $m_S$ [\GeV] \\
\hline
\multirow{6}{*}{Stealth SUSY}  & 250 & \multirow{6}{*}{100 , 90} \bigstrut[t] \\
                                                 & 500 &  \\
                                                 &  800 & \\
                                                 & 1200 & \\
                                                 & 1500 & \\
                                                 & 2000 & \\
\bottomrule
\end{tabular}\label{tab:MCSamples}
\end{table}

Zero-bias data are used to estimate the expected background for the 2MSVx strategy and potential contamination by non-collision background in the 1MSVx+AO strategies. These data are acquired with a special trigger which fires on the bunch crossing that occurs one LHC revolution after a low-threshold calorimeter-based trigger and therefore have a negligible signal contamination. The zero-bias trigger runs throughout ATLAS data taking, so these data are acquired with the same beam conditions present in normal physics data and can be used to study the expected background. Due to the very high output event rate, the zero-bias trigger is prescaled and only a fraction of the total events are recorded. For this reason, the integrated luminosity acquired is much lower than the total collected during 2015 and 2016 data taking and corresponds to 1.1~$\mu$b$^{-1}$ and 12~$\mu$b$^{-1}$ for the two periods, respectively.

Monte Carlo (MC) simulation samples were produced for all models considered in this paper. The masses, summarized in Table~\ref{tab:MCSamples}, were chosen to span the accessible parameter space. For the Stealth SUSY model, the singlino and singlet masses were set to 100~\GeV\ and 90~\GeV, respectively. These values were recommended by the authors of the model as a good representative choice~\cite{Josh1}. The small mass-splitting between the singlino and singlet ensures that the gravitino carries off very little momentum. The \emph{mean proper lifetime} of each sample is tuned to obtain a \emph{mean lab-frame decay length} of 5~m. This choice maximizes the distribution of decays throughout the ATLAS detector volume. The mean proper lifetime used for the generation of the samples is within a range of 0.17--5.55~m, depending on the sample. For each MC sample, 400,000 events are produced. 

Since the analysis is sensitive to a wide range of mean proper lifetimes, and the generation of many samples to cover a broad lifetime range would be extremely CPU-time consuming, a toy MC strategy was adopted to extrapolate the number of expected events to the range of mean proper lifetimes between 0 and 1000\,m. For each LLP in the MC sample a random decay position sampled from an exponential distribution was generated. The physical decay position in the detector was then calculated for each particle using the LLP four-momenta from the simulated MC samples. The overall probability of the event to satisfy the selection criteria was then evaluated from efficiencies to satisfy each selection criterion, parametrized as a function of the LLP decay position.

In order to validate the extrapolation procedure described above, another set of samples for only the scalar boson model with $m_{s} \geq 125~\GeV$ and with fewer (200,000) events was generated. The mean proper lifetime in each of these samples was tuned in order to have a slightly longer mean lab-frame decay length, corresponding to 9 m. The mean proper lifetimes in these MC samples span a range of 0.23--7.20~m, depending on the sample.

All MC samples described above were generated at leading order using \mbox{MG5\_aMC@NLO} 2.2.3~\cite{MadGraph5} interfaced to \mbox{\textsc{Pythia 8.210}}~\cite{PYTHIA8} parton shower model. The A14 set of tuned parameters~\cite{ATL-PHYS-PUB-2014-021} was used together with the NNPDF2.3LO parton distribution function (PDF) set~\cite{Ball:2012cx}. The \textsc{EvtGen 1.2.0} program~\cite{EvtGent} was used for the properties of $b$- and $c$-hadron decays.
The generated events were processed through a full simulation of the ATLAS detector geometry and response~\cite{GEANT4b} using the \textsc{Geant4}~\cite{GEANT4a} toolkit.
The simulation includes multiple \emph{pp} interactions per bunch crossing (pileup), as well as the effect on the detector response due to interactions from bunch crossings before or after the one containing the hard interaction. 
Pileup was simulated with the soft strong-interaction processes of \textsc{Pythia 8.210} using the A2 set of tuned parameters~\cite{ATL-PHYS-PUB-2012-003} and the MSTW2008LO~\cite{MSTW2008} PDF set.
Per-event weights were applied to the simulated events to correct for inaccuracies in the pileup simulation.

\section{Trigger and event reconstruction} \label{sec:trigreco}

Hadronic LLP decays in the MS typically produce narrow, high-multiplicity hadronic showers. Variations in track multiplicity and shower width depend on the mass and boost of the decaying LLP and the final states to which the LLP decays. Dedicated trigger~\cite{MuonRoITrigger} and vertex~\cite{MSVXRECO} algorithms were developed to select and reconstruct displaced decays in the MS. Due to the amount of material in the calorimeter, only decays occurring in or after the last sampling layer of the hadronic calorimeter will generally produce a significant number of hits in the MS and therefore were reconstructed.

\subsection{Reconstruction of prompt hadronic jets and missing transverse momentum} 

Calorimeter jets with a $\ET$ threshold greater than 10~\GeV\ and $|\eta|<4.9$ are constructed at the electromagnetic (EM) energy scale using the \antikt jet algorithm \cite{antkt4} with a radius parameter $R=0.4$ using the FastJet 2.4.3 software package \cite{FASTJET}. A collection of three-dimensional topological clusters of neighboring energy deposits in the calorimeter cells containing a significant energy above a noise threshold~\mbox{\cite{TopoClustRun1,JetEnergyMeas}} provide input to the \antikt algorithm. The calorimeter cell energies are measured at the EM scale, corresponding to the energy deposited by electromagnetically interacting particles. After reconstruction, jets are calibrated using the procedure outlined in Ref.~\cite{JETPERF2}.

The missing transverse momentum, \MET, is defined as the magnitude of the negative vector sum of the transverse momenta of preselected electrons, muons, photons and jets, to which is added an extra term to account for energy deposits that are not associated with any of these selected objects~\cite{ATLAS-CONF-2018-023}. This extra term was calculated from inner detector tracks matched to the PV to make it more resilient to contamination from pileup interactions. For the analysis presented in this paper, electrons, muons and photons were used only in the computation of \MET, and their reconstruction is detailed in Refs.~\cite{ATL-PHYS-PUB-2016-015,PERF-2015-10}. Electrons were required to have a $\pT>10$~\GeV\ and $|\eta|<2.47$ and also pass medium identification requirements~\cite{ATL-PHYS-PUB-2015-041}. Muons were required to have a $\pT>10$~\GeV\ and $|\eta|<2.7$ with a matching track in the ID and pass a medium quality requirement~\cite{PERF-2015-10}. Photons were selected using a tight identification requirement~\cite{ATL-PHYS-PUB-2016-014}. 
Since tracklets (defined in Section~\ref{sec:MSVxReco}) are not used for the \MET calculation, a displaced vertex from a signal event in the MS will contribute to the \MET.

\subsection{Muon RoI Cluster trigger}

The Muon RoI Cluster trigger is a signature-driven trigger that selects candidate events for decays of LLPs particles in the MS: events must contain a cluster of muon RoIs within a $\Delta R=0.4$ cone. The details of the performance and implementation of this trigger can be found in Ref.~\cite{MuonRoITrigger}. The isolation criteria for jets and tracks, discussed in Ref.~\cite{MuonRoITrigger} and used to reduce background punch-through jets, were not applied in the analysis presented in this paper. 
The trigger selects isolated, signal-like events and non-isolated, \mbox{background-like} events. The background-like events were then available to be used in control regions and for data-driven background estimations in signal regions.

\begin{figure}[htbp]
\begin{subfigure}[b]{0.49\textwidth}
\includegraphics[width=\textwidth]{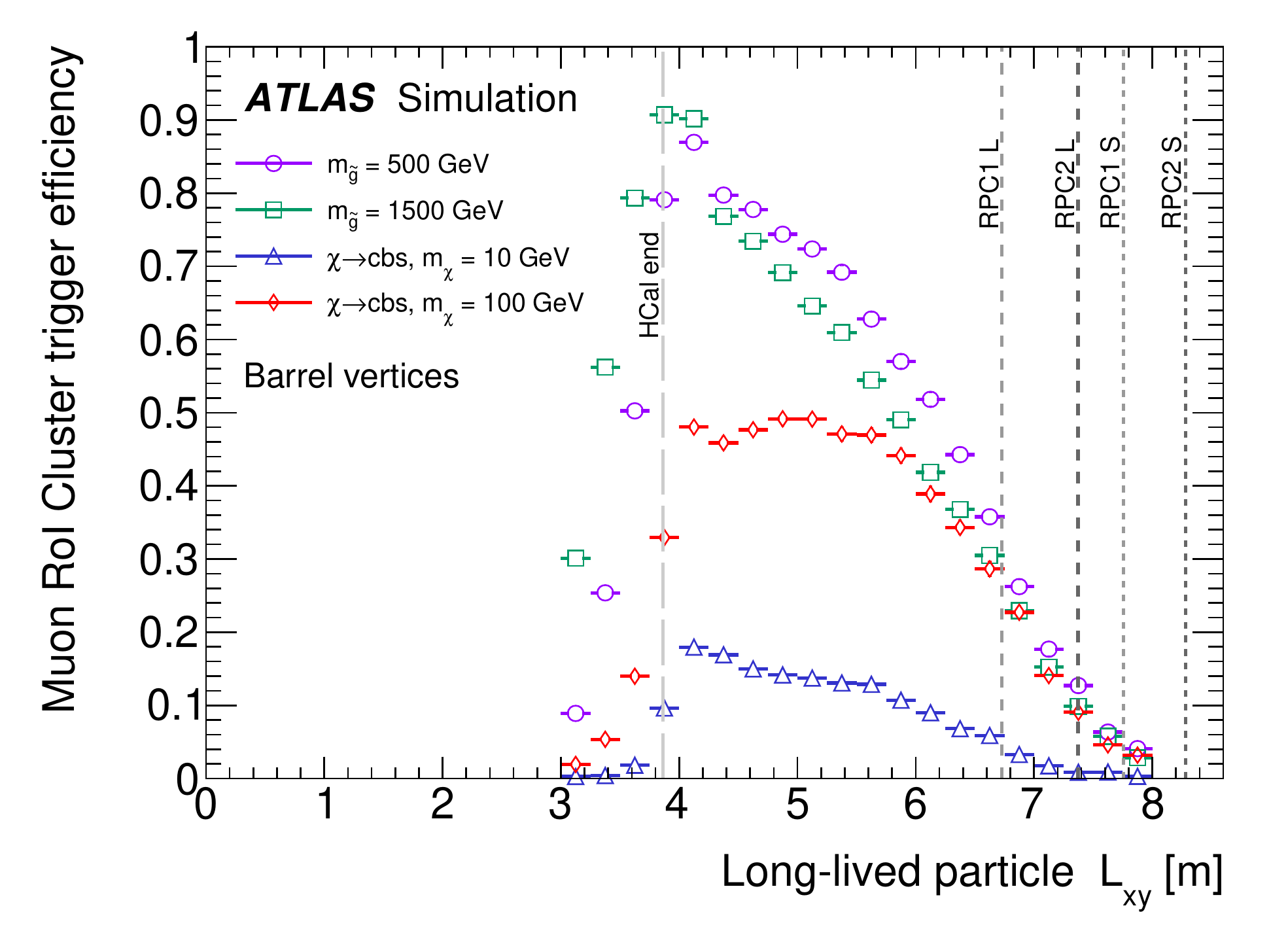}
\caption{}\label{fig:MuonClusterTriggerEffBarrel}
\end{subfigure}
\begin{subfigure}[b]{0.49\textwidth}
\includegraphics[width=\textwidth]{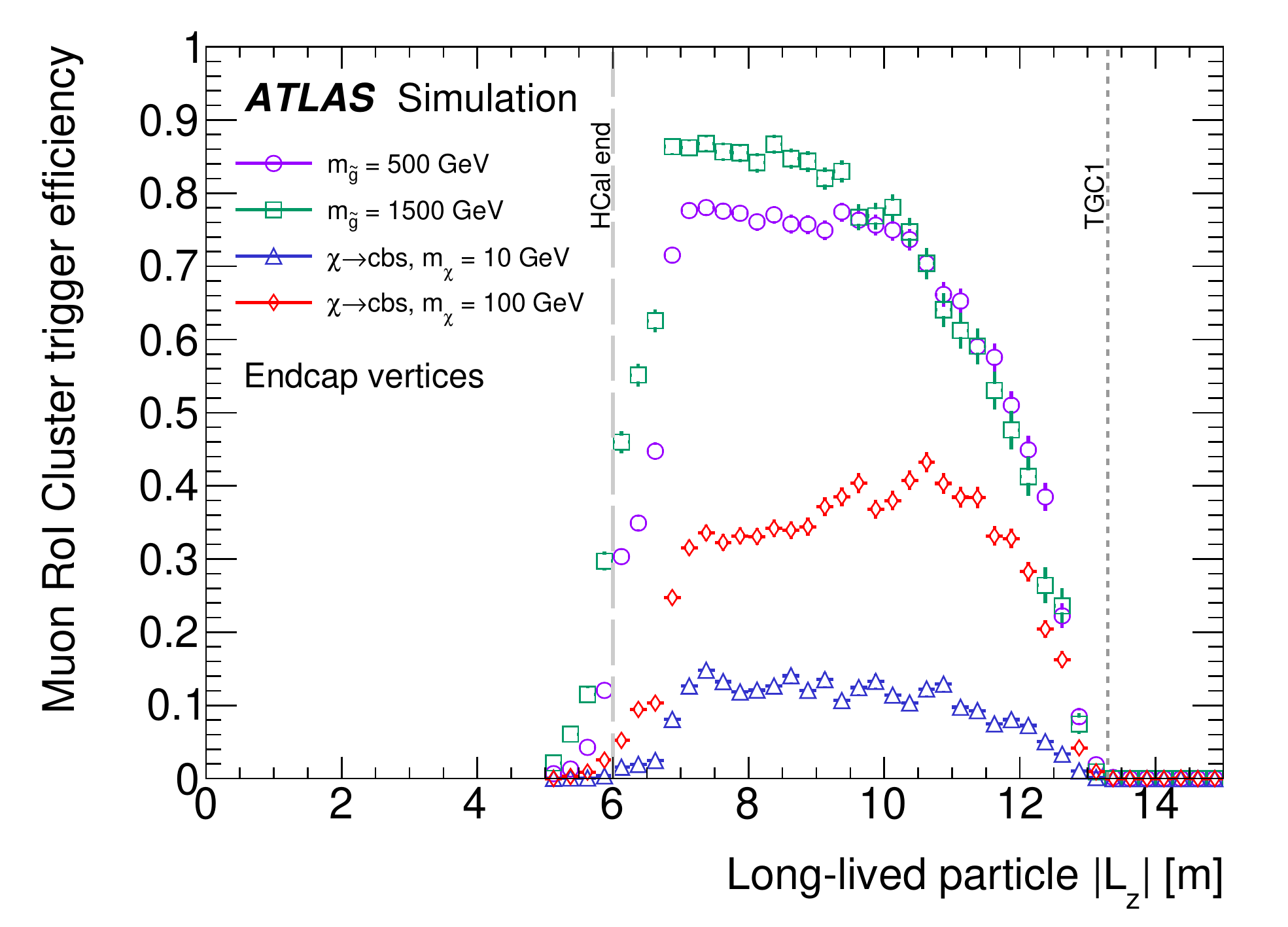}
\caption{}\label{fig:MuonClusterTriggerEffEndcaps}
\end{subfigure}
\caption{Efficiency for the Muon RoI Cluster trigger as a function of the decay position of the LLP for four simulated benchmark samples in the (a) MS barrel and (b) MS \Endcaps. The vertical lines show the relevant detector boundaries, where ``HCal end'' is the outer limit of the hadronic calorimeter, RPC 1/2 represent the first/second layer of RPC chambers, TGC 1 represents the first layer of TGC chambers and L/S indicate whether they are in Large or Small sectors.}
\label{fig:MuonClusterTriggerEff}
\end{figure}

The trigger efficiency, defined as the fraction of LLPs selected by the trigger as a function of the LLP decay position, is shown in Figure~\ref{fig:MuonClusterTriggerEffBarrel} and Figure~\ref{fig:MuonClusterTriggerEffEndcaps} for four MC simulated benchmark samples with LLP decays in the MS barrel and endcap regions, respectively. The efficiency was parameterized as a function of the transverse decay position ($L_{xy}$) in the barrel and the longitudinal decay position ($|L_z|$) in the endcaps. The trigger is efficient for hadronic decays of LLPs that occur anywhere from the outer regions of the HCal to the middle stations of the MS. These efficiencies were obtained from the subset of events with only a single LLP decay in the muon spectrometer in order to ensure that the result of the trigger is due to a single burst of MS activity. The uncertainties shown are statistical only. The relative differences between the efficiencies of the benchmark samples are a result of the different masses of the LLPs, which in turn affect their momenta and consequently the opening angles of the decay products. The trigger efficiency is higher when the LLP decays close to the end of the hadronic calorimeter (barrel: $r \sim 4$~m, endcaps: $z \sim 6$~m) and it decreases substantially as the decay occurs closer to the middle station of the muon spectrometer (barrel: $r \sim 7$~m, endcaps: $z \sim 13$~m). For decays occurring close to the middle station, the charged hadrons and photons (and their EM showers) are not spatially separated and they are overlapping when they traverse the middle stations.

Scale factors were used in order to correct for mismodeling of the L1 muon trigger response in MC simulation and they were calculated by comparing the distributions of the average number of muon RoI clusters within a $\Delta R$ cone of 0.4 around the axis of a punch-through jet in multijet MC and data events. In fact, a high-energy jet has a high probability of punching through into the MS and creating a cluster of muon RoIs that can mimic the behavior of signal events. High-energy jets were selected using a jet trigger with a $\ET$ threshold of 400~\GeV. The scale factor is $1.13\pm0.01$ for the barrel and $1.04\pm0.02$ for the endcaps, and it does not depend on the $\eta$ or the \pT of the jet.

\subsection{Reconstruction of MS vertices}\label{sec:MSVxReco}

A dedicated algorithm~\cite{MSVXRECO}, capable of reconstructing low-momentum tracks in a busy environment, was used to reconstruct the displaced MS vertices used in this search. The algorithm takes advantage of the spatial separation between the two multilayers inside a single MDT chamber. Single-multilayer straight-line segments that contain three or more MDT hits were reconstructed using a minimum $\chi^2$ fit. Segments from multilayer-1 were then matched with those from multilayer-2. The paired set of single-multilayer segments and corresponding track parameters is called a \emph{tracklet}. These tracklets are used to reconstruct the positions of MS vertices. This algorithm was previously  used for both the 7~\TeV~\cite{HVAtlas} and 8~\TeV~\cite{EXOT-2013-12} searches for displaced decays. Detectable decay vertices were located in the region between the outer edge of the HCal and the middle station of muon chambers. Due to the different detector technology (no spatially separated multilayers), the CSC chambers were not used for the MS vertex reconstruction.

\subsubsection{Reconstructed objects for vertex isolation}\label{sec:GVC}

In order to ensure sufficient signal acceptance and background rejection, a set of \emph{vertex isolation criteria} for ID tracks and calorimeter jets was established in order to assist in determining whether or not a vertex is consistent with a displaced hadronic decay.  

For track isolation, two separate criteria were used: one for high-\pT tracks which considers tracks with $\pT > 5$~\GeV, and one for large multiplicities of low-\pT tracks which used the \pT vector sum of all tracks associated with the primary vertex (PV) with $\pT > 400$~\MeV\ in a $\Delta R$ cone of 0.2 around the MS vertex axis\footnote{The MS vertex axis is defined with respect to the detector coordinate system.}. The two different isolations stem from the fact that some jets have most of their energy in a single hadron, while others can consist of multiple low-\pT tracks. 

For the 2MSVx and 1MSVx+Jets strategies, all the jets considered for isolation must meet \emph{jet quality criteria}. Jets must satisfy $\ET>30$~\GeV\ and $\LogCalRatio<0.5$. The value $\LogCalRatio$ quantifies the fraction of energy of the jet that is deposited in the HCal ($E_{\textrm{HAD}}$) with respect to the energy deposited in the ECal ($E_{\textrm{EM}}$). This requirement ensures that vertices originating from LLPs that decay near the outer edge of the hadronic calorimeter and also have significant MS activity were not rejected. In addition, in order to reduce the probability that a signal vertex fails to meet the isolation criteria due to pileup jets that do not have sufficient energy to create an MS vertex, jets with $20 < \ET < 60$~\GeV\ were required to be matched to the PV using a jet vertex tagger (JVT) discriminant~\cite{PERF-2014-03}. Standard jet quality criteria\,\cite{ATLAS-CONF-2015-029} were not enforced because jets that do not fulfill these requirements can also produce a background MS vertex.

For the 1MSVx+\MET strategy a looser selection on the jets used for isolation was applied. The reason for this is that most of the background that enters the signal region of this strategy is generated by events where a jet satisfying the jet quality criteria is almost back-to-back with an MS vertex created by another jet that does not fulfill the jet quality criteria, but has enough energy to punch through into the muon spectrometer. Studies performed in data showed that there could be standard or pileup jets with a measured energy down to 15~\GeV\ that can create a vertex in the MS. These jets would not be used to compute the isolation since they do not pass the $\ET$ selection requirement present in jet quality criteria and the associated vertex would be incorrectly considered as a signal candidate. For these reasons, all jets above 15~\GeV\ were considered in the isolation computation for the 1MSVx+\MET strategy. The looser selection on the jets used for isolation has negligible impact on the signal efficiency, according to simulation; in all the samples considered for the results presented in this paper there are no events rejected because of a jet below 20~\GeV\ entering the isolation.

The same type of jet events could also affect the 2MSVx and 1MSVx+Jets strategies. For the former the jet quality criteria have no impact on the background estimation, while for the latter the effect is negligible because the additional selection on the energy of the two prompt jets strongly reduces this background contribution. 

Vertex isolation criteria were optimized separately for each analysis strategy described in Section~\ref{sec:strategy}, and they are described in detail in Sections~\ref{sec:2msvx} and \ref{sec:1msvx}.

\section{Baseline event selection} \label{sec:baseselection}

A common baseline selection was applied to the events considered in the three strategies described in Table~\ref{tab:topo}. Events were required to pass the Muon RoI Cluster trigger and contain a PV with at least two tracks with $\pT > 400$~\MeV. The vertex with the largest sum of the squares of the transverse momenta of all tracks associated with the vertex was chosen as the PV. 
This PV selection has no impact on the signal efficiency. In simulation, the selected PV corresponds to the signal interaction in about 95--99\% of the cases, depending on the sample; even though the LLPs are invisible in the ID, the resonance (scalar, Higgs boson) is produced with a significant \pT.

An MS vertex due to a displaced decay typically has many more hits than an MS vertex from background; consequently a minimum number of MDT ($n_{\textrm{MDT}}$) and RPC/TGC ($n_{\textrm{RPC}}$/$n_{\textrm{TGC}}$) hits was required. The number of MDT hits was counted in the MDT chambers that have their center within $\Delta \phi = 0.6$ and $\Delta \eta = 0.6$ of the vertex $(\eta, \phi)$ position. The number of RPC or TGC hits is the sum of hits that are within $\Delta R = 0.6$ of the vertex position. A requirement on the maximum number of MDT hits was also applied to remove background events caused by coherent noise bursts in the MDT chambers. In addition to reducing the background, the minimum required number of RPC/TGC hits helps to further reject these noisy events, because a noise burst in the MDT system is not expected to be coherent with one in the muon trigger system.

A displaced decay that occurs in the transition region between MS barrel and endcaps results in hits in both regions. Vertex reconstruction was performed separately in the barrel and endcaps, and only the barrel (endcap) hits were used in the barrel (endcap) vertex reconstruction algorithm. Therefore, any vertices reconstructed from either of the two algorithms have fewer hits, as they were reconstructed from a subset of the total hits. The result is a decrease in the reconstruction efficiency, and this also occasionally results in two vertices being reconstructed from a single LLP decay.
Therefore, the MS vertices with pseudorapidity, $|\eta_{\textrm{vx}}|$, between 0.8 and 1.3 were not considered in the analysis. This has a negligible impact on the signal efficiency, since the average MS vertex efficiency in this region is less than 2\%.

Background studies performed for the 1MSVx+AO strategies using data showed that in the transition region between the barrel and the endcap hadronic calorimeters, $0.7 < |\eta_{\textrm{vx}}| < 1.2$, the probability of having a jet that does not fulfill the minimal selection criteria for being considered for isolation and that punches through into the MS is much higher than in other regions of the detector. This region overlaps the already excluded MS transition region, except for  $0.7 < |\eta_{\textrm{vx}}| < 0.8$. The fraction of signal events removed was very small compared to the gain obtained by removing punch-through jet background that could affect the single-vertex analysis; therefore, vertices reconstructed in the MS region 
$0.7 < |\eta_{\textrm{vx}}| < 0.8$ were not considered either. 

Table~\ref{tab:ABCDCutflow} summarizes the baseline criteria used to select ``good'' MS vertices.

\begin{table}[!h]
\centering
\caption{Summary of the baseline criteria used for the analysis presented in this paper. All selection criteria are also applied to signal MC events when determining the number of expected signal events in the dataset.}
\begin{tabular}{p{6 cm}p{6 cm}}
\toprule
 \multicolumn{2}{l}{Event passes Muon RoI Cluster trigger} \\ 
 \midrule
 \multicolumn{2}{l}{Event has a PV with at least two tracks with $\pT > 400$~\MeV} \\
 \midrule
 \multicolumn{2}{l}{Event has at least one MS vertex} \\
 \midrule
 \multicolumn{2}{l}{MS vertex matched to triggering muon RoI cluster ($\Delta R(\text{vertex}, \text{cluster}) < 0.4$).} \\ 
 \multicolumn{2}{l}{For 2MSVx strategy: in the case of 2 muon RoI clusters, the second vertex should be} \\
  \multicolumn{2}{l}{matched to the second cluster.} \\
  \midrule
 \multicolumn{2}{l}{300 $\le n_{\textrm{MDT}} <$ 3000} \\                
\bottomrule
\end{tabular}

\vspace{0.1 cm}

\begin{tabular}{p{6 cm}p{6 cm}}
 \toprule
 \emph{\hspace{1.5 cm} Barrel} & \emph{\hspace{1.5 cm} Endcaps} \\
  \midrule
 MS vertex with $|\eta_{\textrm{vx}}| < 0.7$ & MS vertex with $1.3 < |\eta_{\textrm{vx}}| < 2.5$ \\
 $n_{\textrm{RPC}} \geq 250$ & $n_{\textrm{TGC}} \geq 250$ \\                               
\bottomrule
\end{tabular}
\label{tab:ABCDCutflow}
\end{table}
\section{Two-MS-vertex search} \label{sec:2msvx}

The two-MS-vertex strategy is designed to be sensitive to models where the LLP is pair-produced and decays hadronically between the outer region of the HCal and the middle station of the MS. 
Requiring two displaced vertices significantly reduces the expected background.
In addition, background from punch-through jets was further reduced using the isolation criteria described in Section~\ref{sec:GVC}. 

Residual background can arise from collision or non-collision processes and cannot be accurately simulated. Thus, data-driven methods were used to estimate the expected background, which also avoids systematic uncertainties due to the use of simulated events.

\subsection{Event selection}\label{sec:2MSVxSelection}

In order to improve the rejection of background from punch-through jets, the isolation criteria using the reconstructed objects described in Section~\ref{sec:GVC} were optimized for all the benchmark samples considered in this paper by comparing signal with multijet simulated events. The isolation criteria used for the 2MSVx strategy are summarized in Table~\ref{tab:vertexIso}, where $\Delta R$ is defined as the angular distance between the direction of the tracks or jets and the vertex axis. An MS vertex with tracks and/or jets satisfying these criteria was not considered in the analysis.

\begin{table}[h!]
\centering
\caption{Summary of the isolation criteria used to select signal events for the 2MSVx strategy in the barrel and endcaps regions. $\Delta R$ is defined as the angular distance between the direction of the tracks or jets and the vertex axis. MS vertices satisfying these criteria were not considered in the analysis.}
\begin{tabular}{lcc}
\toprule
\multicolumn{1}{c}{Isolation requirements for 2MSVx strategy} & Barrel & Endcaps \\
\midrule
High-\pT track isolation, ($p_T > 5$~\GeV) & $\Delta R < 0.3$ & $\Delta R <$ 0.6 \\
Low-\pT track isolation, $(\Sigma \pT(\Delta R<0.2))$ & $\Sigma \pT < 10$~\GeV\ & $\Sigma \pT < 10$~\GeV\ \\
Jet isolation & $\Delta R < 0.3$ & $\Delta R < 0.6$  \\
\bottomrule
\end{tabular}
\label{tab:vertexIso}
\end{table}

At least two isolated MS vertices must be present in the events. One MS vertex must be matched to the trigger-level muon RoI cluster [$\Delta R(\textrm{cluster, vertex}) < 0.4$]. If there were two distinct clusters, each MS vertex must be matched to one cluster. To ensure that the two MS vertices and/or two muon RoI trigger clusters do not come from the \emph{same} background activity, the two vertices were required to be separated by at least $\Delta R = 1.0$, which has minimal impact on the overall signal acceptance.

\subsection{MS vertex efficiency}\label{sec:2MSVxEff}

\begin{figure}[htbp]
\begin{subfigure}[b]{0.49\textwidth}
	\includegraphics[width=\textwidth]{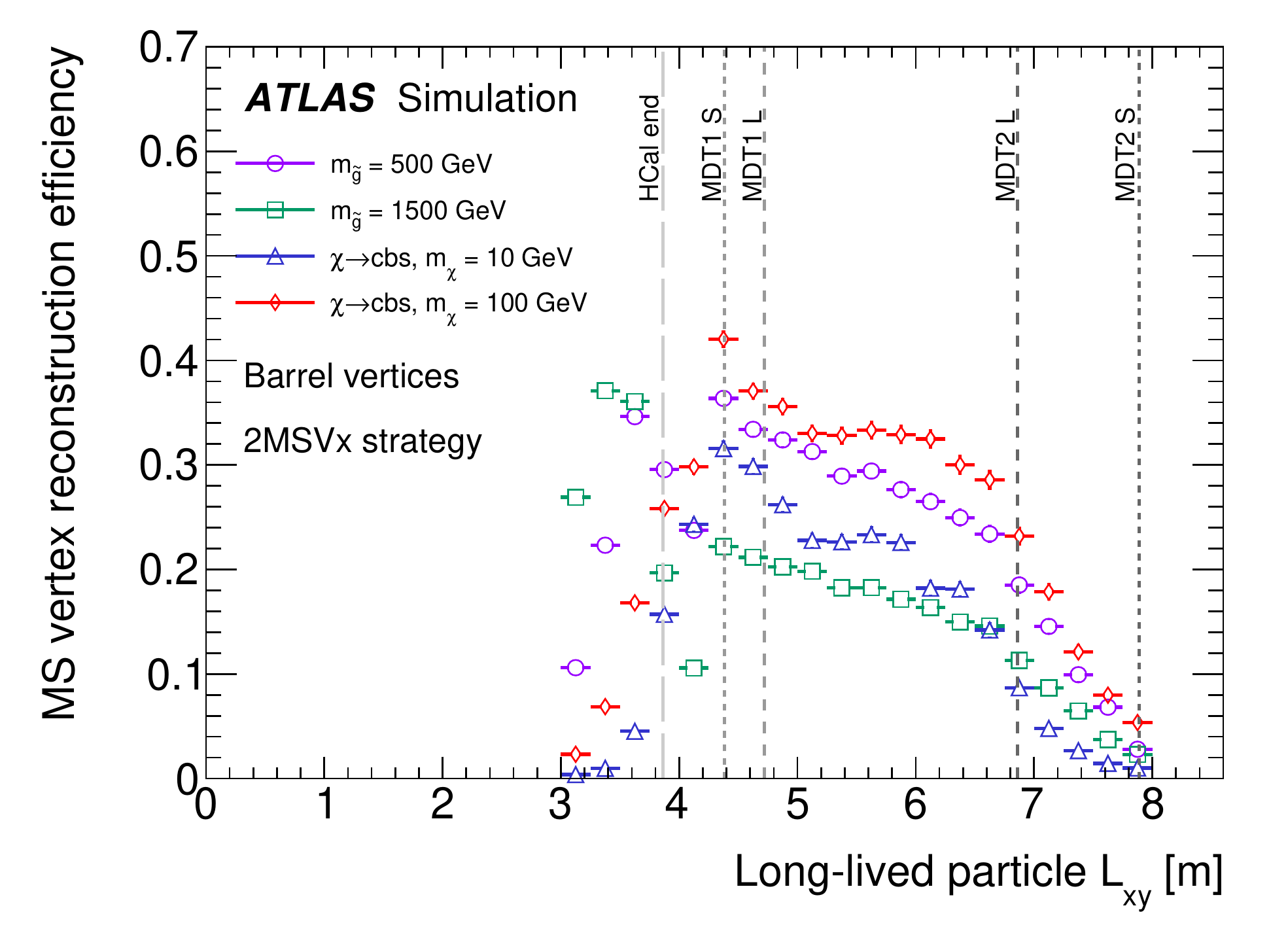}
	\caption{}  
	\label{fig:MSVertexEffBarrel}
	\end{subfigure}
\begin{subfigure}[b]{0.49\textwidth}
	\includegraphics[width=\textwidth]{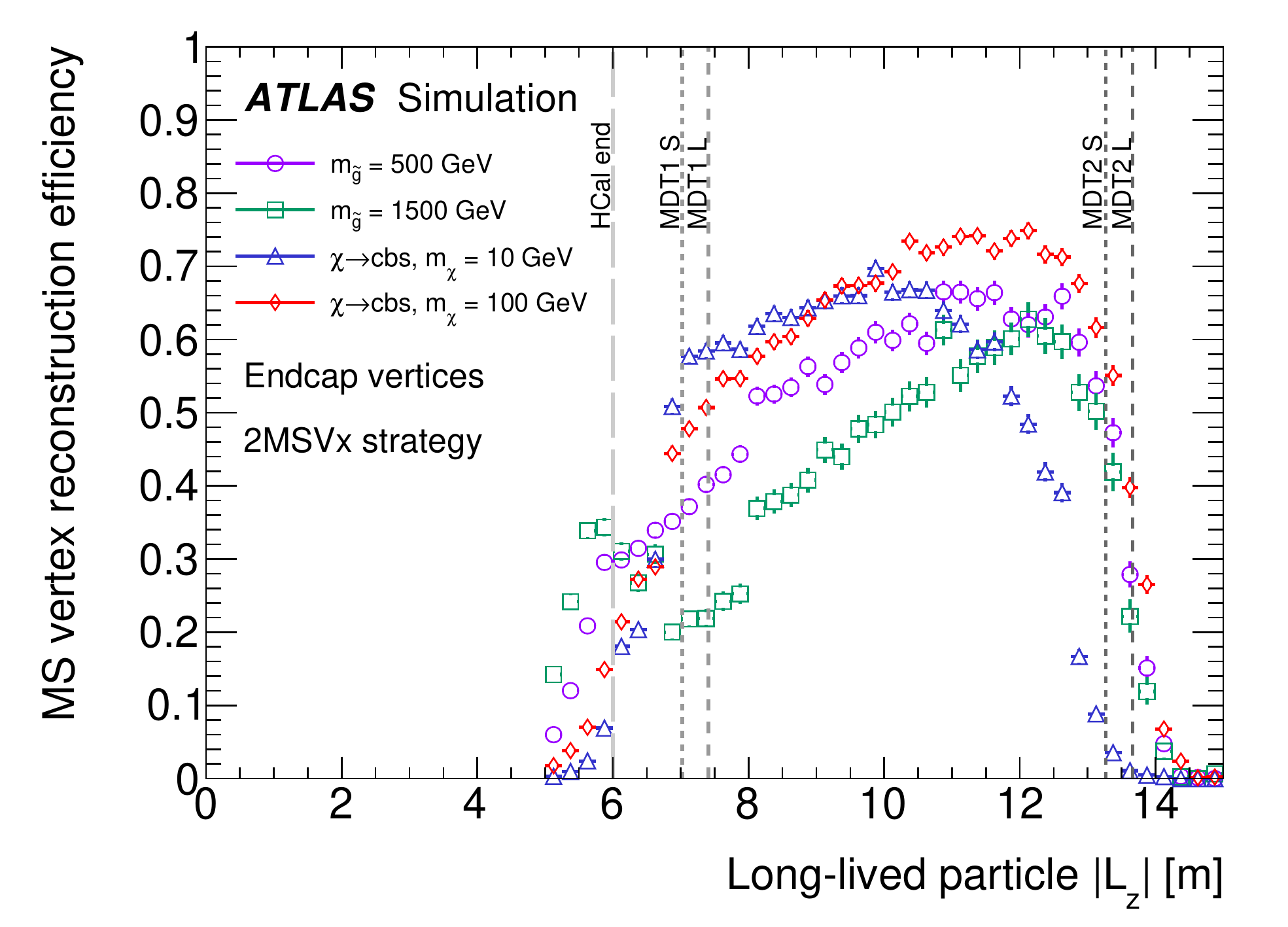}
	\caption{}
	\label{fig:MSVertexEffEndcap}
	\end{subfigure}
	\caption{Efficiency to reconstruct an MS vertex for some of the Stealth SUSY and baryogenesis benchmark samples, for vertices that pass the baseline event selection and satisfy the vertex isolation criteria (no trigger selection is applied). (a): Barrel MS vertex reconstruction efficiency as a function of the transverse decay position of the LLP. (b): Endcap MS vertex reconstruction efficiency as a function of the longitudinal decay position of the LLP relative to the center of the detector. The vertical lines show the relevant detector boundaries, where ``HCal end'' is the outer limit of the hadronic calorimeter, MDT 1/2 represent the first/second layer of MDT chambers and L/S indicate whether they are in Large or Small sectors.}
	\label{fig:MSVertexEffs}
\end{figure}

The efficiency for vertex reconstruction is defined as the fraction of simulated LLP decays in the MS fiducial volume which match a reconstructed vertex passing the baseline event selection and satisfying the vertex isolation criteria~\cite{MSVXRECO}. A reconstructed vertex is considered matched to a displaced decay if the vertex is within $\Delta R = 0.4$  of the simulated decay position. The MS vertex efficiency was parameterized as a function of the transverse ($L_{xy}$) and longitudinal ($|L_z|$) LLP decay position in the barrel and endcaps, respectively. Figure~\ref{fig:MSVertexEffBarrel} shows the efficiency for reconstructing a vertex in the MS barrel for a selection of benchmark samples. Figure~\ref{fig:MSVertexEffEndcap} shows the efficiency for reconstructing a vertex in the MS endcaps.

The MS barrel vertex reconstruction efficiency is 30--40\% near the outer edge of the hadronic calorimeter ($r\approx4$\,m) and it decreases substantially as the decay occurs closer to the middle station ($r\approx7$\,m). The decrease occurs because the charged hadrons and photons are not spatially separated and overlap when they traverse the middle station. This results in a reduction of the efficiencies for tracklet reconstruction and, consequently, vertex reconstruction. The efficiency for reconstructing vertices in the MS endcaps reaches 70\% for higher-mass benchmark models. Because there is no magnetic field in the region in which endcap tracklets are reconstructed, the vertex reconstruction algorithm does not have the constraints on charge and momentum that are present in the barrel. Consequently, the vertex reconstruction in the endcaps is more efficient for signal, but also less robust in rejecting background events. More details are provided in Ref.~\cite{MSVXRECO}.

\subsection{Background estimation}\label{sec:Bkg2MSVx}

To estimate the expected background for the 2MSVx strategy it is necessary to quantify the frequency with which the MS vertex algorithm reconstructs isolated vertices for non-signal events. This number can be calculated from data using events with one isolated MS vertex which pass either the Muon RoI Cluster trigger or a zero-bias trigger. 
The expected background with two isolated MS vertices is calculated as follows:
\begin{equation*}
N_\textrm{2Vx} = N^\textrm{1cl} \cdot P^\textrm{Vx}_\textrm{noMStrig} + 
                            N^\textrm{2cl}_\textrm{1UMBcl} \cdot P^\textrm{Vx}_\textrm{Bcl} + 
                            N^\textrm{2cl}_\textrm{1UMEcl} \cdot P^\textrm{Vx}_\textrm{Ecl} \,\,\,\,\, .
\end{equation*}
The events selected by the Muon RoI Cluster trigger and containing only one MS vertex are separated into those containing only one cluster of muon RoIs ($N^\textrm{1cl}$), and those containing two muon RoI clusters, where only one cluster is matched to the reconstructed MS vertex and the other is unmatched in the barrel or endcaps ($N^\textrm{2cl}_\textrm{1UMBcl}$,$ N^\textrm{2cl}_\textrm{1UMEcl}$). The term $P^\textrm{Vx}_\textrm{noMStrig}$ is the probability of finding a vertex in events not selected by the Muon RoI Cluster trigger. This probability is determined from zero-bias events by dividing the number of good, isolated MS vertices not passing the muon RoI Cluster trigger by the total number of zero-bias events that satisfy standard event quality criteria. The terms $P^\textrm{Vx}_\textrm{Bcl}$ and $P^\textrm{Vx}_\textrm{Ecl}$ are the probabilities for finding an MS vertex given a muon RoI cluster in the barrel and endcaps, respectively. 
Since zero events are observed with two trigger clusters and one vertex, the contribution of the $N^\textrm{2cl}$ terms is negligible. 	

Therefore, the number of two-MS-vertex events can be calculated as $N_\textrm{2Vx} = N^\textrm{1cl} \cdot P^\textrm{Vx}_\textrm{noMStrig}$.

Six good isolated MS vertices were found in 35,673,956 zero-bias triggered events, while 159,816 events with one isolated muon RoI cluster matched to a vertex were selected in the 2015 and 2016 datasets. The zero-bias sample has no overlap with the Muon RoI Cluster triggered events and contains zero events with more than one MS vertex. Contamination from signal events would result in overestimation of the probabilities and resulting background rates. The probability $P^\textrm{Vx}_\textrm{noMStrig}$ is thus estimated to be $6/35,673,956 = (1.7 \pm 0.7) \times10^{-7}$, where the uncertainty is statistical only. Therefore, the expected number of background events with one trigger cluster and two vertices is evaluated as $(159,816 \pm 400)  \cdot (1.7 \pm 0.7) \times 10^{-7} = 0.027 \pm 0.011$, where the uncertainty is statistical only.

\section{Single-MS-vertex search} \label{sec:1msvx}

For models with two LLPs, the probability of having both LLPs decay inside the detector decreases for mean lab-frame decay lengths greater than $\sim 5$~m. Thus, extending sensitivity to shorter and longer proper lifetimes for a given model also requires a strategy of using only one reconstructed displaced decay~\cite{LLPDataDriven}. In the regime of long lifetimes the single-vertex analysis in the MS has unique sensitivity compared to other displaced searches, although it is affected by higher levels of background. For a search with only one displaced object, a background determination method similar to the two-vertex search does not work since the ensemble of events with one isolated vertex, used to estimate the background, already contains the signal region for the 1MSVx+AO strategies. Instead, non-isolated vertices are used in a data-driven method to estimate the expected number of isolated fake vertices. 

The following sections describe the event selection and background estimation for the 1MSVx+Jets and 1MSVx+\MET strategies. The events considered in these searches must satisfy the baseline selection criteria summarized in Section~\ref{sec:baseselection}.

\subsection{Event selection}\label{sec:SigSelection1MSVx}

Two separate signal selections are used for the two topologies that are considered in the single-MS-vertex search.

\subsubsection{1MSVx+Jets strategy}\label{sec:SigSelMSVxJets}

\begin{figure}[h!]
	\centering 
    \begin{subfigure}[b]{0.48\textwidth} 
	\includegraphics[width=\textwidth]{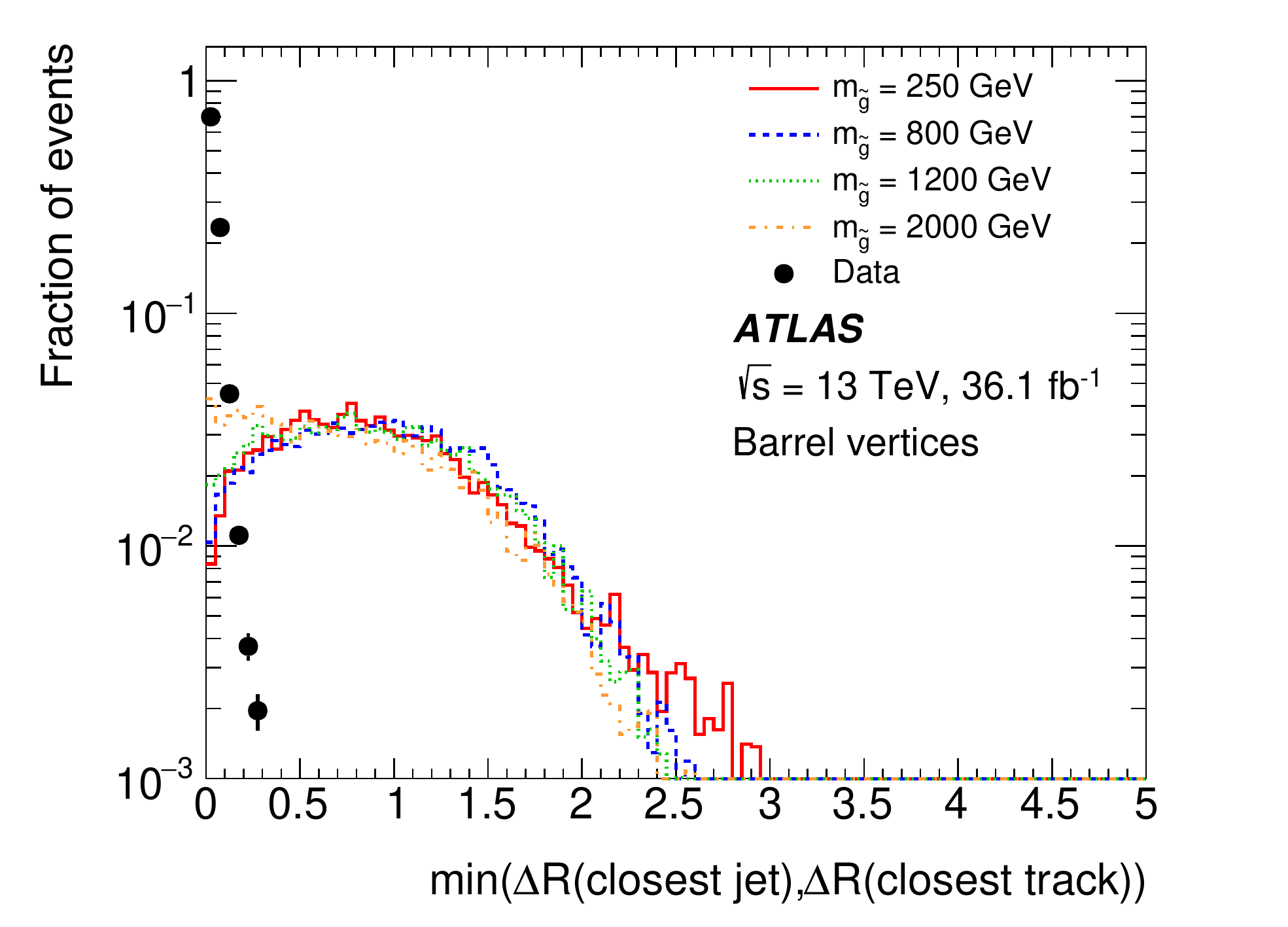}
	\caption{}\label{fig:IsoVar_stealth_a}
    \end{subfigure}
     \begin{subfigure}[b]{0.48\textwidth}  \vspace{-0.1 cm}
     	\includegraphics[width=\textwidth]{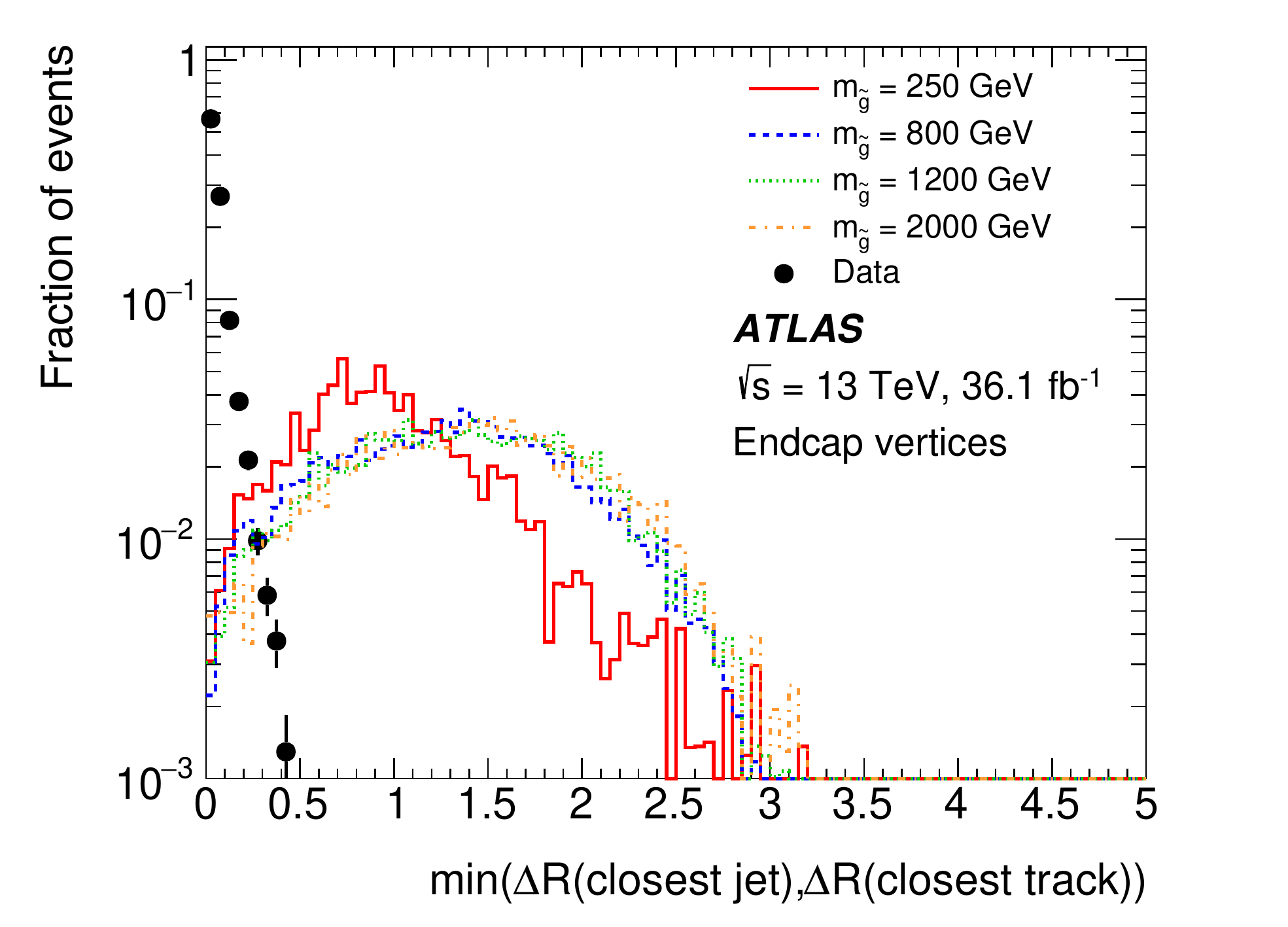}
	\caption{}\label{fig:IsoVar_stealth_b}
    \end{subfigure} \\ 
      \caption{Distributions of the isolation variable used to select signal events for the 1MSVx+Jet strategy, for the (a) barrel and (b) endcaps. The black points are data collected in 2015 and 2016, while the dashed and solid lines show the distributions for four of the MC Stealth SUSY signal samples. The events in the plots satisfy the baseline selection criteria described in Section~\ref{sec:baseselection}. Distributions are normalized to unity.}
    \label{fig:IsoVar_stealth}
   \end{figure}

The main criterion that is used to distinguish a signal MS vertex from background is its degree of isolation as described in Section~\ref{sec:2MSVxSelection}. 
To characterize the degree of isolation with a single value, the variable $\Delta R_{\text{min}}= \mbox{min}\big(\Delta R(\mbox{vertex, closest jet}), \Delta R(\mbox{vertex, closest track})\big)$ was defined. Figures~\ref{fig:IsoVar_stealth_a} and \ref{fig:IsoVar_stealth_b} show the distributions of the isolation variable used for 1MSVx+Jets events for data and some of the MC benchmark samples for barrel and endcap vertices, respectively. 

\begin{figure}[h!]
	\centering 
     \begin{subfigure}[b]{0.48\textwidth}
	\includegraphics[width=\textwidth]{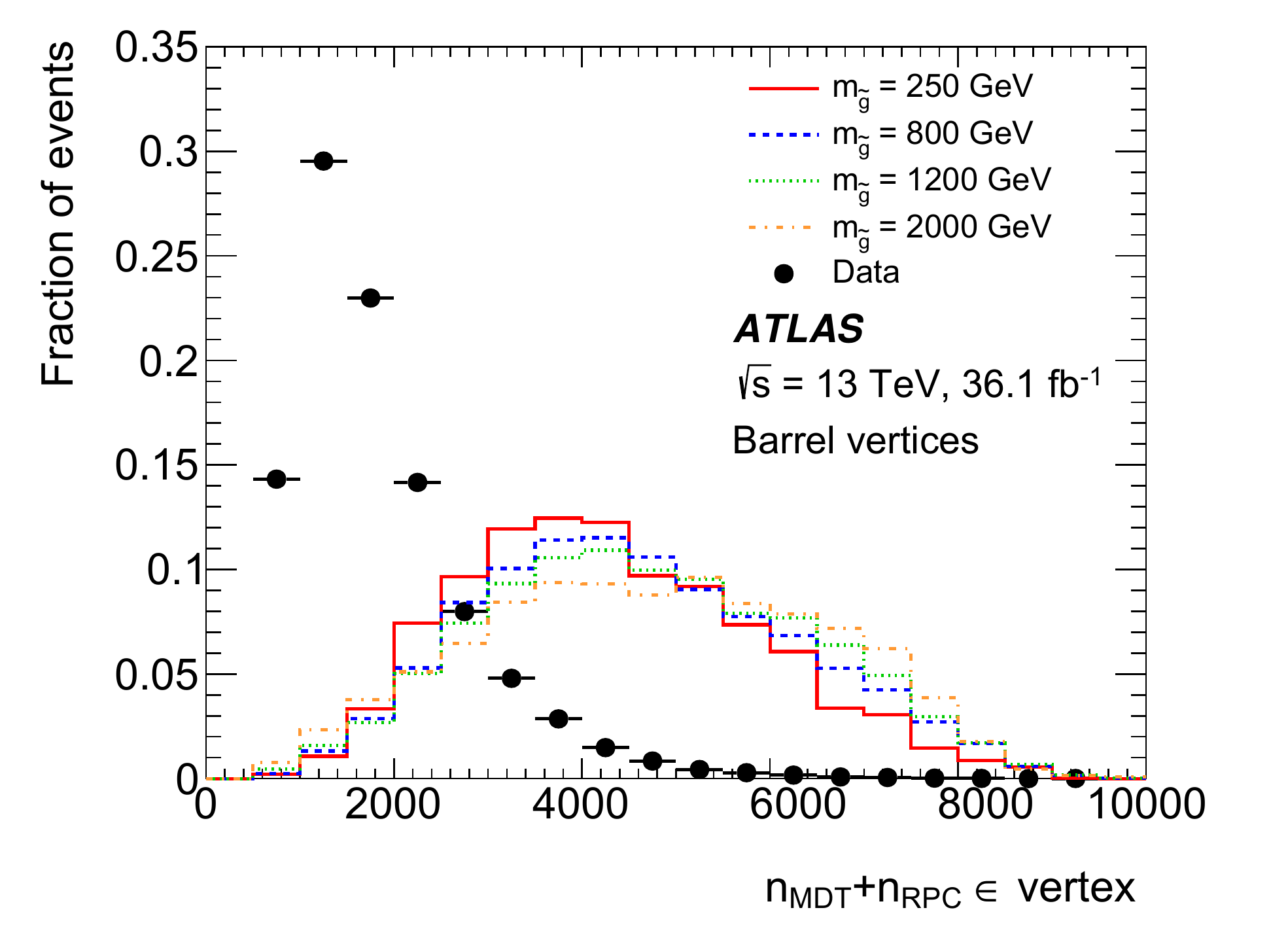}
	\caption{}\label{fig:HitsVar_stealth_a}
    \end{subfigure}
    \begin{subfigure}[b]{0.48\textwidth}
	\includegraphics[width=\textwidth]{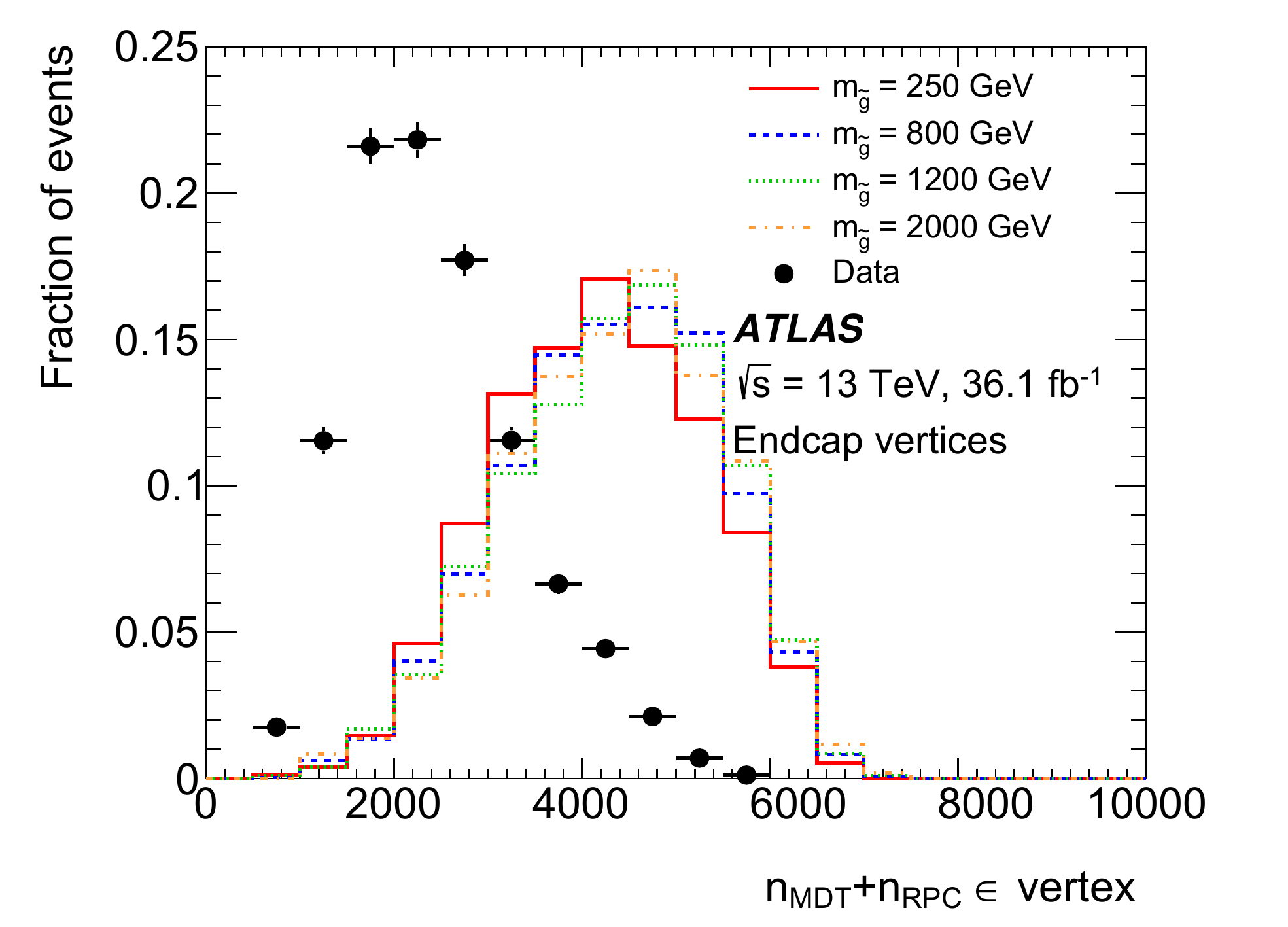}
	\caption{}\label{fig:HitsVar_stealth_b}
    \end{subfigure}  \\ \vspace{-0.2 cm}
      \caption{Distributions of the number of MS hits associated with the displaced vertex, used to select signal events for the 1MSVx+Jet strategy, for the (a) barrel and (b) endcaps. The black points are data collected in 2015 and 2016, while the dashed and solid lines show the distributions for four of the MC Stealth SUSY signal samples. The events in the plots satisfy the baseline selection criteria described in Section~\ref{sec:baseselection}. Distributions are normalized to unity.}
    \label{fig:HitsVar_stealth}
   \end{figure}

Another signal selection variable is the sum of the number of MDT hits and trigger hits (RPC and TGC in the barrel and endcaps, respectively) in a cone around an MS vertex, since a signal event is expected to leave more hits than a background one. Figures \ref{fig:HitsVar_stealth_a} and \ref{fig:HitsVar_stealth_b} present the distributions of the number of MS hits variable used for the 1MSVx+Jets strategy for data collected during 2015 and 2016 and for some of the MC benchmark samples for barrel and endcap vertices, respectively.

Moreover, the two prompt jets produced by the gluino decays can be used to improve the signal selection for the Stealth SUSY analysis. The second-highest (subleading) jet $E_\text{T}$ is generally above 150~\GeV, though applying this requirement results in some loss of signal efficiency in the lowest-mass gluino sample ($m_{\tilde{g}}=250$~\GeV). For events with a barrel MS vertex the $E_\text{T}$ of the leading and subleading jets was required to be above 150~\GeV, while for events with an endcap MS vertex, a tighter requirement of 250~\GeV\ was chosen due to the higher levels of background. Since the isolation variable depends on the $\Delta R$ between a jet and the vertex, jets chosen for this selection must have $\Delta R(\text{jet, vertex}) > 0.7$. This prevents the selection of a sample containing punch-through jets that leave a vertex in the MS. In signal events, this requirement has minimal effect, since jets and vertices originate from different particles and thus tend to be well-separated. In data events, the main background is from multijet production, and thus if a jet is near a vertex it is generally well within $\Delta R = 0.7$. The selection on the two prompt jets described above is used to define two regions: one signal-dominated, and one background-dominated used to validate the data-driven background estimation.

The signal selection for the 1MSVx+Jets strategy was optimized by examining the signal acceptance and background rejection using data from the background-dominated region defined by the \ET values of the two prompt jets. The minimum values required for isolation $\Delta R_{\text{min}}$ are 0.3 and 0.4 for the barrel and endcaps, respectively; a minimum of 2000 MDT+RPC hits in the barrel and 2500 MDT+TGC hits in the endcaps is required. Table~\ref{tab:1MSVxJetsSigSelection} summarizes the signal selection for the 1MSVx+Jets strategy.

\begin{table}[!h]
\centering
\caption{Summary of the signal selection for the 1MSVx+Jets strategy. An MS vertex satisfying these criteria is selected.}
\begin{tabular}{p{15.45 cm}}
\toprule
Event passes baseline selection \\ 
\bottomrule
\end{tabular}
\vspace{0.1 cm}
\begin{tabular}{p{7.5 cm}p{7.5 cm}}
 \toprule
 \emph{\hspace{1.5 cm} Barrel} & \emph{\hspace{1.5 cm} Endcaps} \\
  \midrule
 $n_{\textrm{MDT}} + n_{\textrm{RPC}} > 2000$   & $n_{\textrm{MDT}} + n_{\textrm{TGC}} > 2500$ \\  
  $\Delta R_{\text{min}} > 0.3$                                & $\Delta R_{\text{min}} > 0.4$ \\ 
  Two jets with $E_{\textrm{T}} > 150$~\GeV\ , $\Delta R(\text{jet,Vx}) > 0.7$ & Two jets with $E_{\textrm{T}} > 250$~\GeV\ , $\Delta R(\text{jet,Vx}) > 0.7$ \\                                                                   
 \midrule                     
\bottomrule
\end{tabular}
\label{tab:1MSVxJetsSigSelection}
\end{table}

\subsubsection{1MSVx+\MET strategy}\label{sec:SigSelMSVxMET}

The one-vertex searches for the scalar portal model with $m_{\varPhi}=125$~\GeV\ and the Higgs portal baryogenesis model are particularly challenging due to the absence of any distinctive associated objects produced with the LLPs, such as the two prompt jets in the Stealth SUSY model.

All the events used for the 1MSVx+\MET strategy are required to have a $\MET > 30$~\GeV, and in the endcaps, only vertices with at least five tracklets are considered, which further reduces the higher level of background present in this region. 

\begin{figure}[h]
	\centering 
    	\begin{subfigure}[b]{0.48\textwidth}
	    \includegraphics[width=\textwidth]{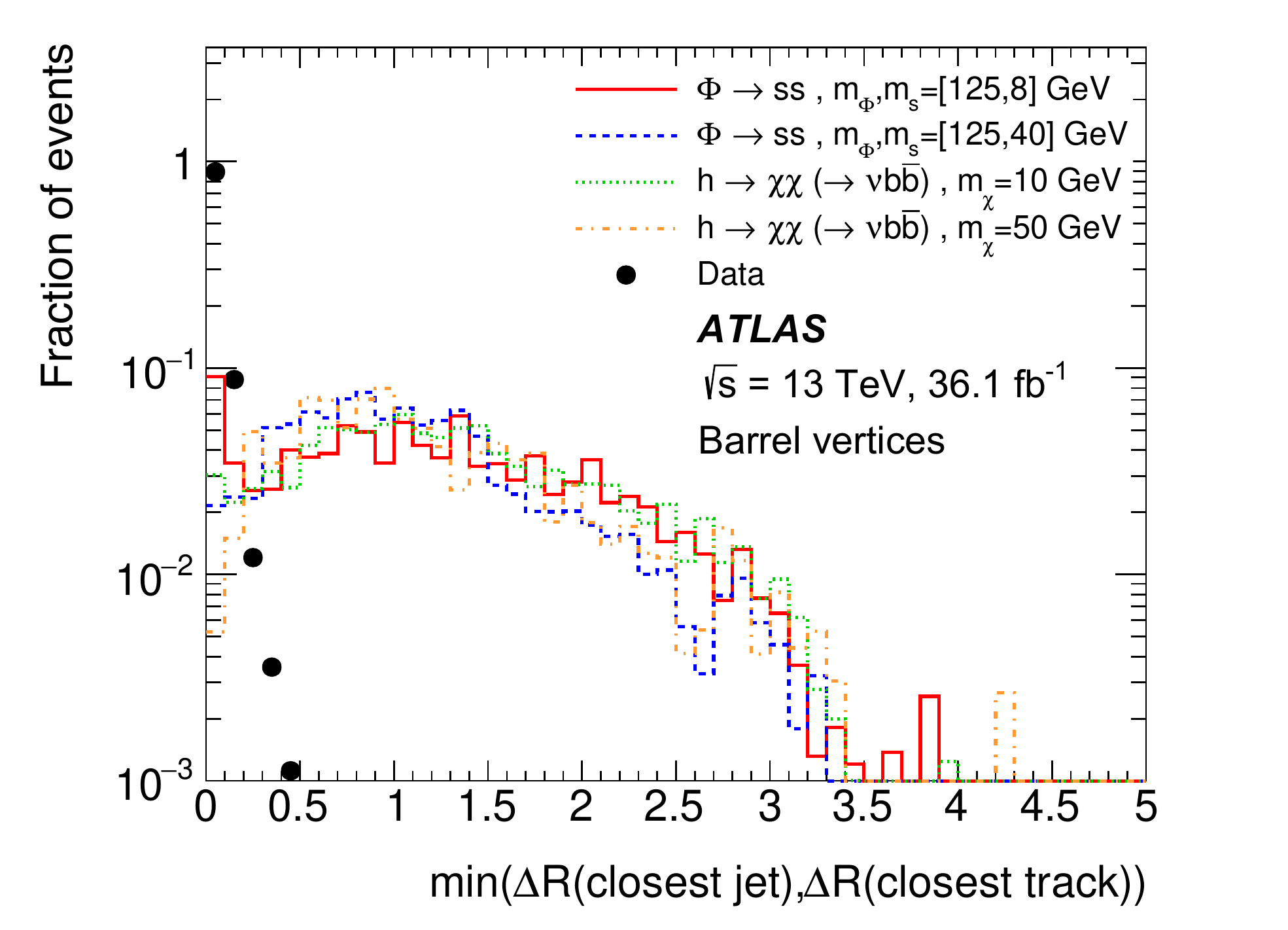}
   	\caption{}
	\label{fig:IsoVar_hss_a}
        \end{subfigure}
    	    \begin{subfigure}[b]{0.48\textwidth}
	\includegraphics[width=\textwidth]{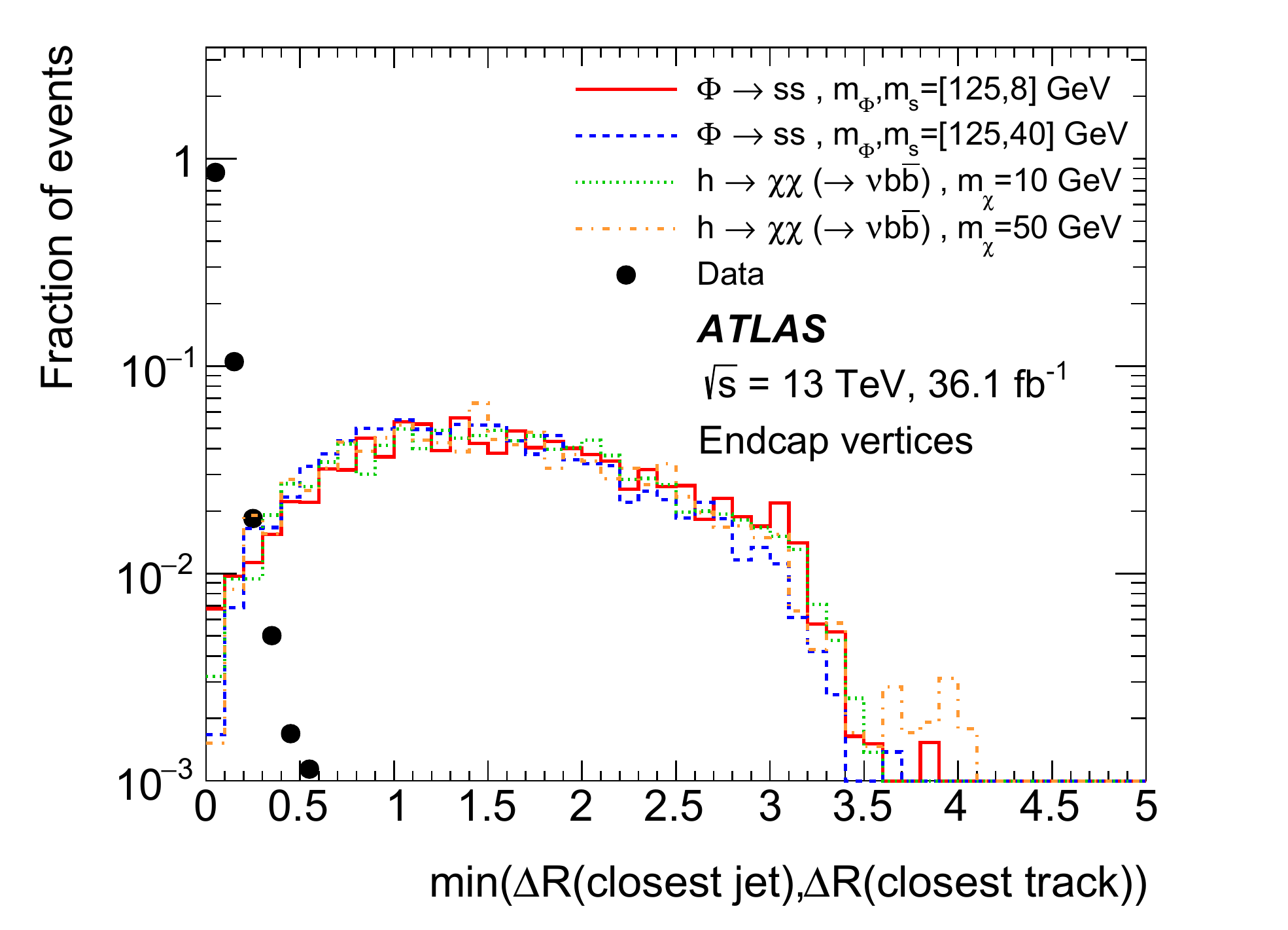}
		\caption{}
	\label{fig:IsoVar_hss_b}
    \end{subfigure}
      \caption{Distributions of the isolation variable used to select signal events for the 1MSVx+\MET strategy, for the (a) barrel and (b) endcaps. The black points are data collected in 2015 and 2016, while the solid lines show the distributions for four MC signal samples. The events in the plots satisfy the baseline selection criteria described in Section~\ref{sec:baseselection} and have $\MET > 30$~\GeV. Distributions are normalized to unity.}
    \label{fig:IsoVar_hss}
   \end{figure}

The same isolation variable, $\Delta R_{\text{min}}$, defined for the 1MSVx+Jets strategy, is also used to select signal events for the 1MSVx+\MET strategy, although for the latter the isolation criteria are computed while placing a looser selection on the jets, as described in Section~\ref{sec:GVC}. Figures~\ref{fig:IsoVar_hss_a} and \ref{fig:IsoVar_hss_b} show the distributions of the isolation variable used for the 1MSVx+\MET strategy for data and some of the MC benchmark samples for barrel and endcap vertices, respectively.

\begin{figure}[h!]
	\centering
      \begin{subfigure}[b]{0.48\textwidth}
	\includegraphics[width=\textwidth]{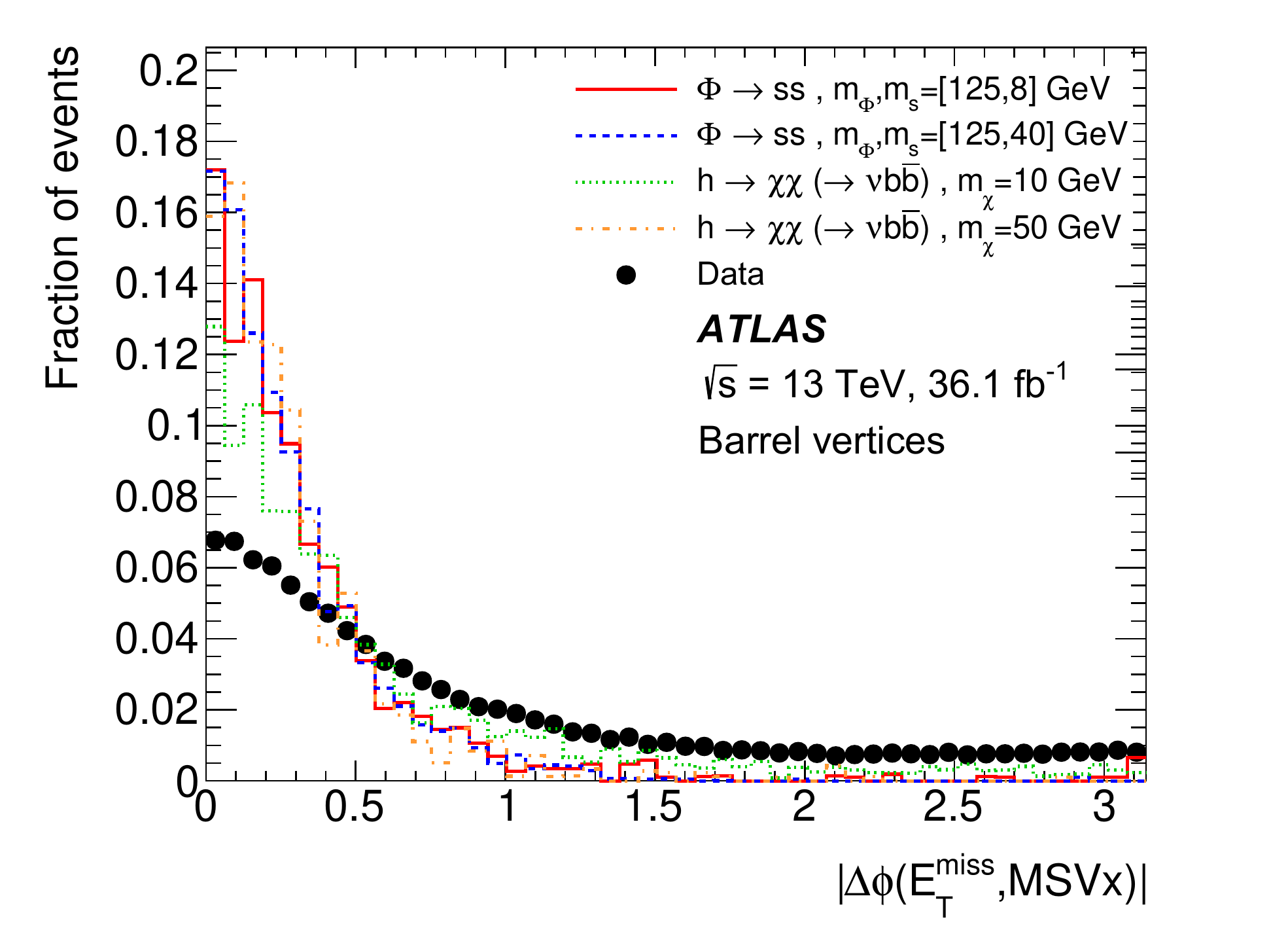}
	\caption{}
	\label{fig:DeltaphiVar_hss_a}
    \end{subfigure}
    	    \begin{subfigure}[b]{0.48\textwidth}
	\includegraphics[width=\textwidth]{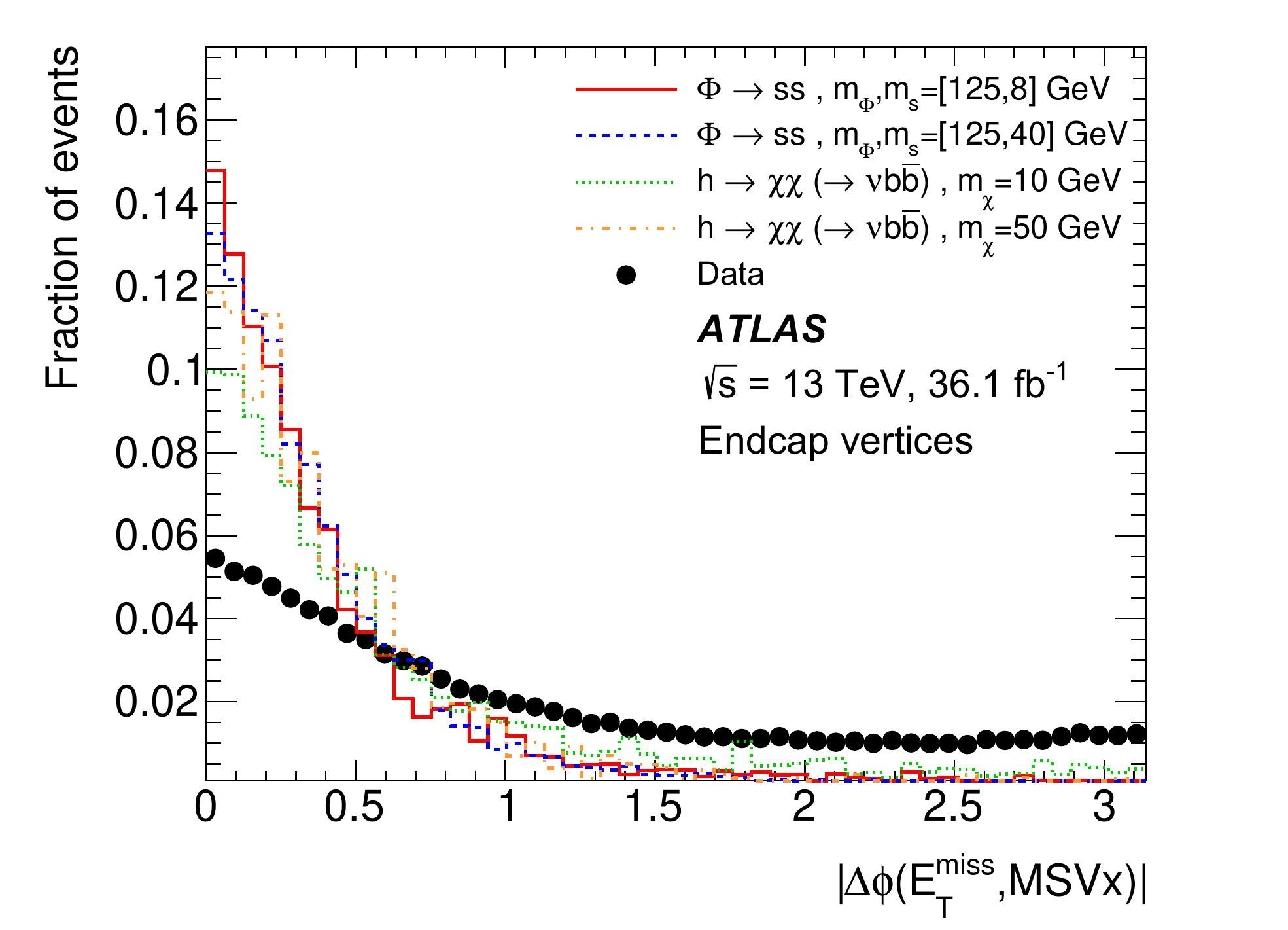}
		\caption{}
	\label{fig:DeltaphiVar_hss_b}
    \end{subfigure} \\ \vspace{-0.2 cm}
      \caption{Distributions of the angle in the transverse plane between the \MET and the displaced vertex $|\Delta\phi(\textrm{\MET,MSVx})|$ used to select signal events for the 1MSVx+\MET strategy, for the (a) barrel and (b) endcaps. The black points are data collected in 2015 and 2016, while the dashed and solid lines show the distributions for four MC signal samples. The events in the plots satisfy the baseline selection criteria described in Section~\ref{sec:baseselection} and have $\MET > 30$~\GeV. Distributions are normalized to unity.}
    \label{fig:DeltaphiVar_hss}
   \end{figure}

The angle in the transverse plane between the \MET vector and the direction of the displaced vertex measured from the origin of the detector coordinate system, $|\Delta\phi(\textrm{\MET,MSVx})|$, was also used to distinguish signal from background because for signal vertices the \MET vector tends to be aligned with the direction of the displaced vertex. Figures \ref{fig:DeltaphiVar_hss_a} and \ref{fig:DeltaphiVar_hss_b} report the distributions of the $|\Delta\phi(\textrm{\MET,MSVx})|$ variable, used to select signal events for the 1MSVx+\MET strategy, for data and some of the MC benchmark samples for barrel and endcap vertices, respectively. 

The number of MDT hits and trigger hits (RPC and TGC hits in the barrel and endcaps, respectively) in a $\Delta R$ cone around the MS vertex was used to define one signal-dominated region and one background-dominated region that was used to validate the data-driven background estimation. The selection requirements that define the two regions were optimized in order to ensure a sufficient signal acceptance and background rejection. For events with a barrel MS vertex the number of hits ($n_{\textrm{MDT}}$+$n_{\textrm{RPC}}$) is required to be greater than 1200, while for events with an endcap MS vertex, the number of hits ($n_{\textrm{MDT}}$+$n_{\textrm{TGC}}$) must exceed 1500.

The signal region selection for the 1MSVx+\MET strategy was optimized using signal acceptance versus background rejection estimated from data in the background-dominated region. The selection requires values of at least 0.8 for isolation $\Delta R_{\text{min}}$ and 1.2 for $|\Delta\phi(\textrm{\MET,MSVx})|$, for both the barrel and endcaps. Due to the higher levels of background, the selection requirement imposed on the isolation is stricter than the ones used for the 1MSVx+Jets strategy. Table~\ref{tab:1MSVxMETSigSelection} summarizes the signal selection for the 1MSVx+\MET strategy.

\begin{table}[!h]
\centering
\caption{Summary of the signal selection for the 1MSVx+\MET strategy. An MS vertex satisfying these criteria is selected.}
\begin{tabular}{p{10.45 cm}}
\toprule
 Event passes baseline selection \\ 
 \midrule
 \MET > 30~\GeV\ \\     
 \midrule
 $|\Delta\phi(\textrm{\MET,MSVx})| < 1.2$ \\
 \midrule
 $\Delta R_{\text{min}} > 0.8$ \\
\bottomrule
\end{tabular}
\vspace{0.1 cm}
\begin{tabular}{p{5 cm}p{5 cm}}
 \toprule
 \emph{\hspace{1.5 cm} Barrel} & \emph{\hspace{1.5 cm} Endcaps} \\
  \midrule
 $n_{\textrm{MDT}} + n_{\textrm{RPC}} > 1200$ & $n_{\textrm{MDT}} + n_{\textrm{TGC}} > 1500$ \\  
                                                                             & $n_{\textrm{tracklets}} \geq 5$ \\  
 \midrule                     
\bottomrule
\end{tabular}
\label{tab:1MSVxMETSigSelection}
\end{table}

\subsubsection{MS vertex efficiency}\label{sec:MSVxEff}

\begin{figure}[htbp]
\begin{subfigure}[b]{0.49\textwidth}
	\includegraphics[width=\textwidth]{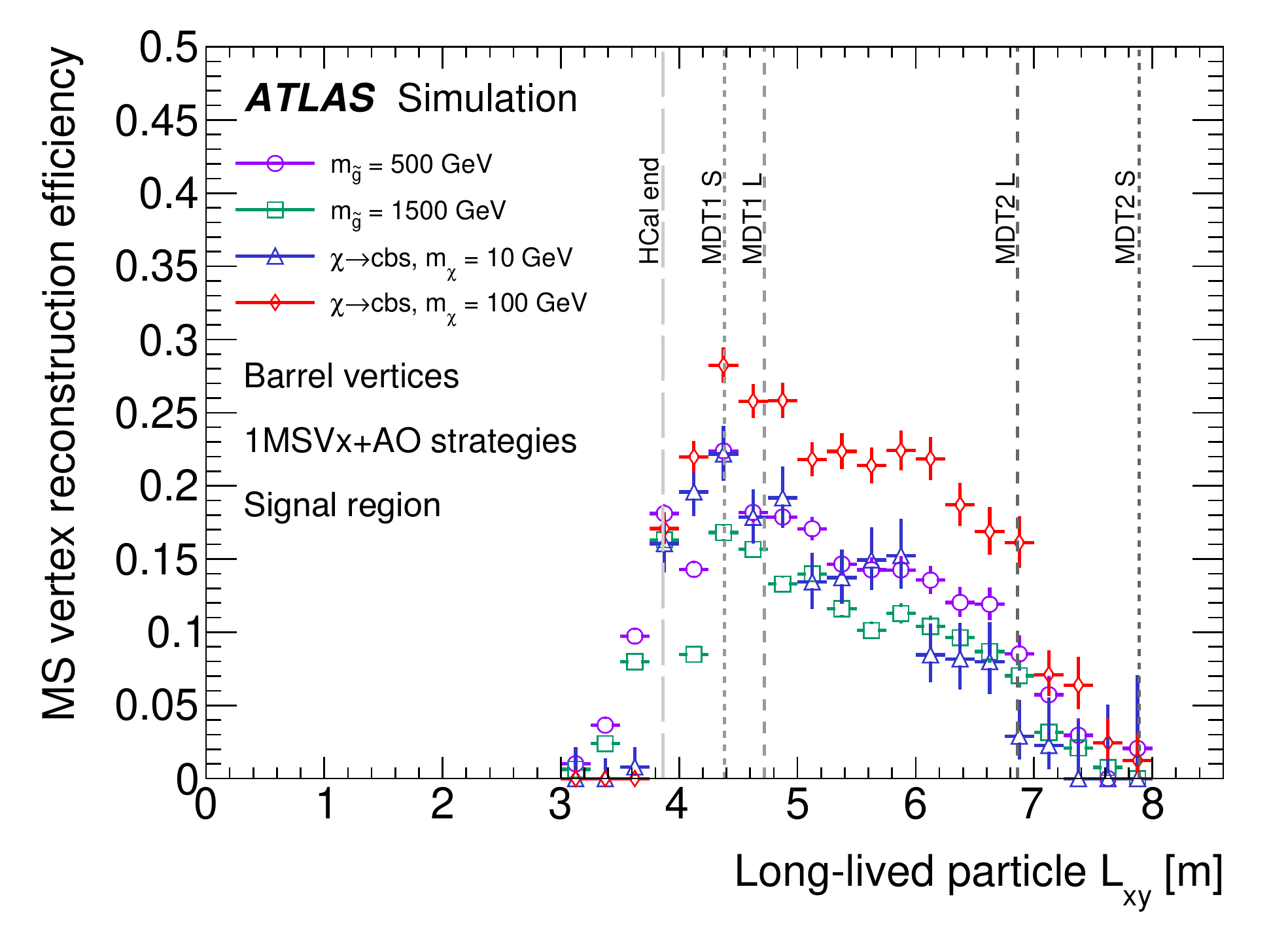}
	\caption{}  
	\label{fig:MSVertexEffBarrel_1MSVx}
	\end{subfigure}
\begin{subfigure}[b]{0.49\textwidth}
	\includegraphics[width=\textwidth]{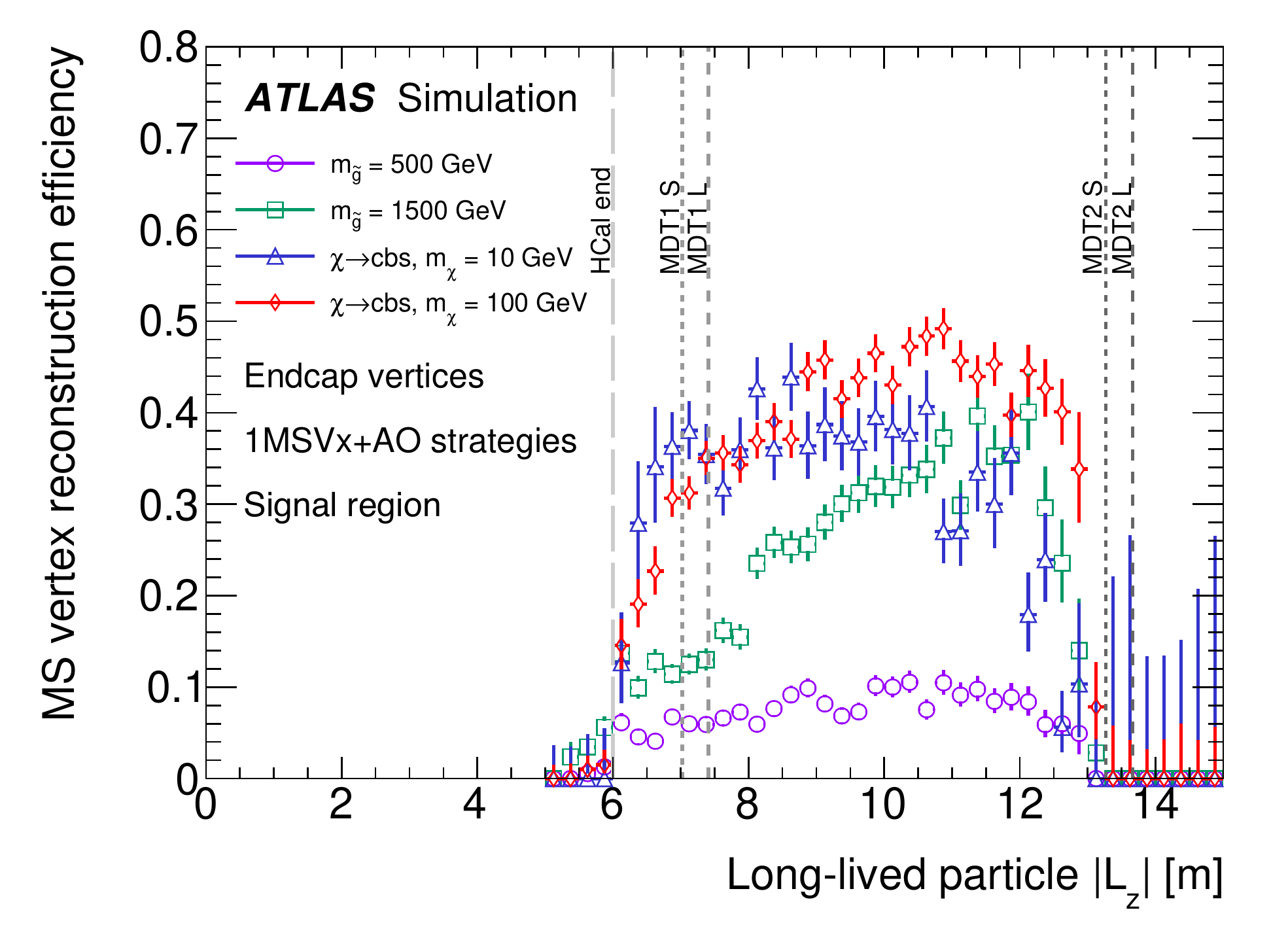}
	\caption{}
	\label{fig:MSVertexEffEndcap_1MSVx}
	\end{subfigure}
	\caption{Efficiency to reconstruct an MS vertex for some of the Stealth SUSY and baryogenesis benchmark samples, for vertices that satisfy the signal selection criteria for the 1MSVx+AO strategies (no trigger selection is applied). (a): Barrel MS vertex reconstruction efficiency as a function of the transverse decay position of the LLP. (b): Endcap MS vertex reconstruction efficiency as a function of the longitudinal decay position of the LLP. The vertical lines show the relevant detector boundaries.}
	\label{fig:MSVertexEffs_1MSVx}
\end{figure}

The efficiency for vertex reconstruction is defined as the fraction of simulated LLP decays in the MS fiducial volume which match a reconstructed vertex satisfying the signal selection criteria. A reconstructed vertex is considered matched to a displaced decay if the vertex is within $\Delta R=0.4$ of the simulated decay position. Figures~\ref{fig:MSVertexEffBarrel_1MSVx} and \ref{fig:MSVertexEffEndcap_1MSVx} show the efficiency for reconstructing a vertex for a selection of benchmark samples in the MS barrel and endcaps, respectively. Vertices selected for the Stealth SUSY and baryogenesis benchmark samples must satisfy the signal selection criteria described in Sections~\ref{sec:SigSelMSVxJets} and \ref{sec:SigSelMSVxMET}, respectively (no trigger selection is applied). The behavior of the MS vertex reconstruction efficiency as a function of the LLP proper lifetime is similar to that shown in Figure~\ref{fig:MSVertexEffs}, although the efficiency values are different due to the different event selection. The MS barrel vertex reconstruction efficiency is 15--25\% near the outer edge of the hadronic calorimeter ($r\approx4$\,m) and it substantially decreases as the decay occurs closer to the middle station ($r\approx7$\,m). The efficiency for reconstructing vertices in the MS endcaps reaches 40\% for baryogenesis and Stealth SUSY high-mass gluino benchmark models. The lower efficiency for the Stealth SUSY sample with $m_{\tilde{g}}=500$~\GeV\ is due to the selection requirement on the $\ET$ values of the two prompt jets, which is not optimal for the lower masses.

\subsection{Background estimation}\label{sec:ABCD}

\begin{figure}[htbp]
\begin{subfigure}[b]{0.49\textwidth}
	\includegraphics[width=\textwidth]{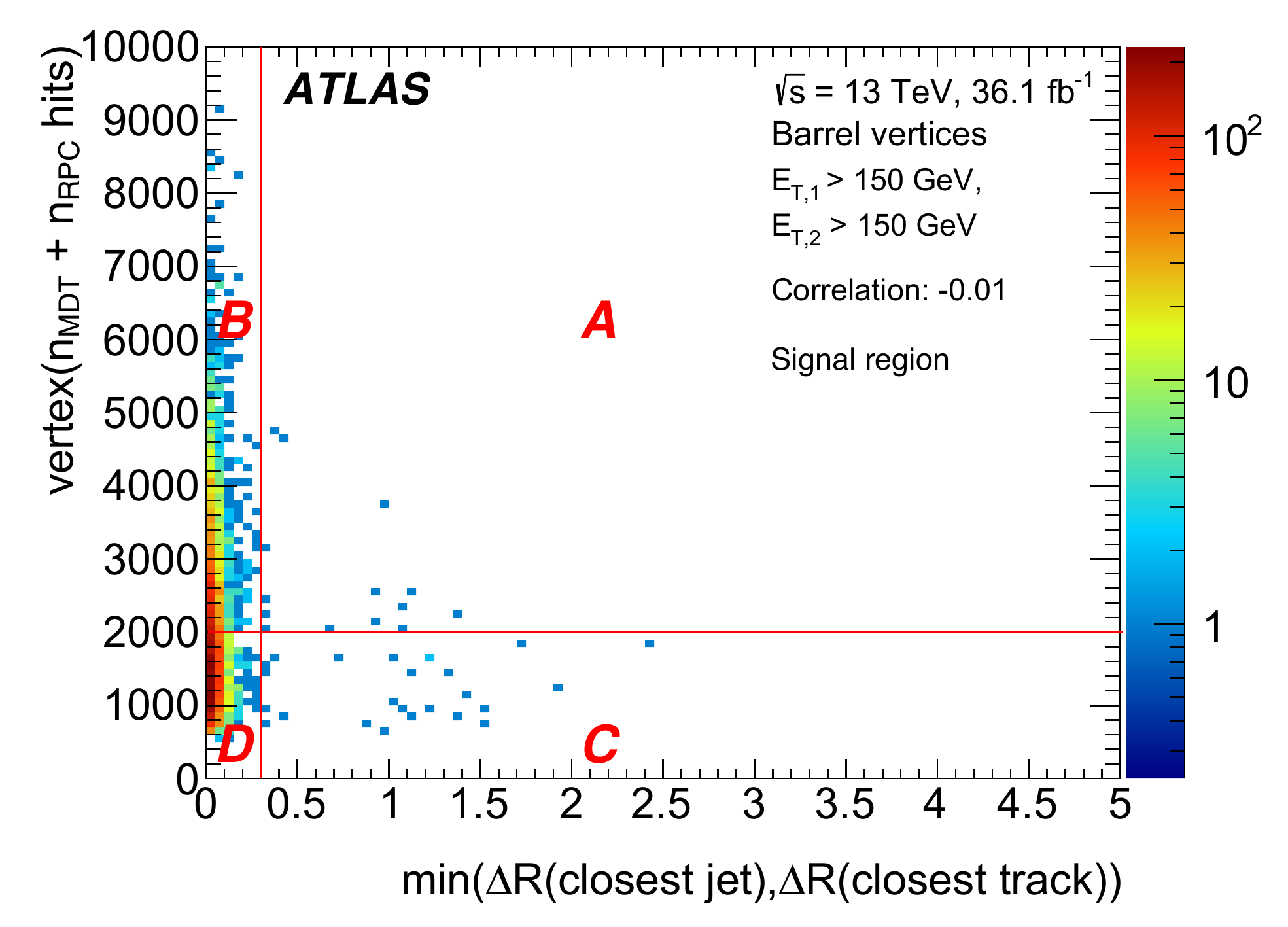}
	\caption{}  
	\label{fig:ABCDplane_b_1MSVxJets}
	\end{subfigure}
\begin{subfigure}[b]{0.49\textwidth}
	\includegraphics[width=\textwidth]{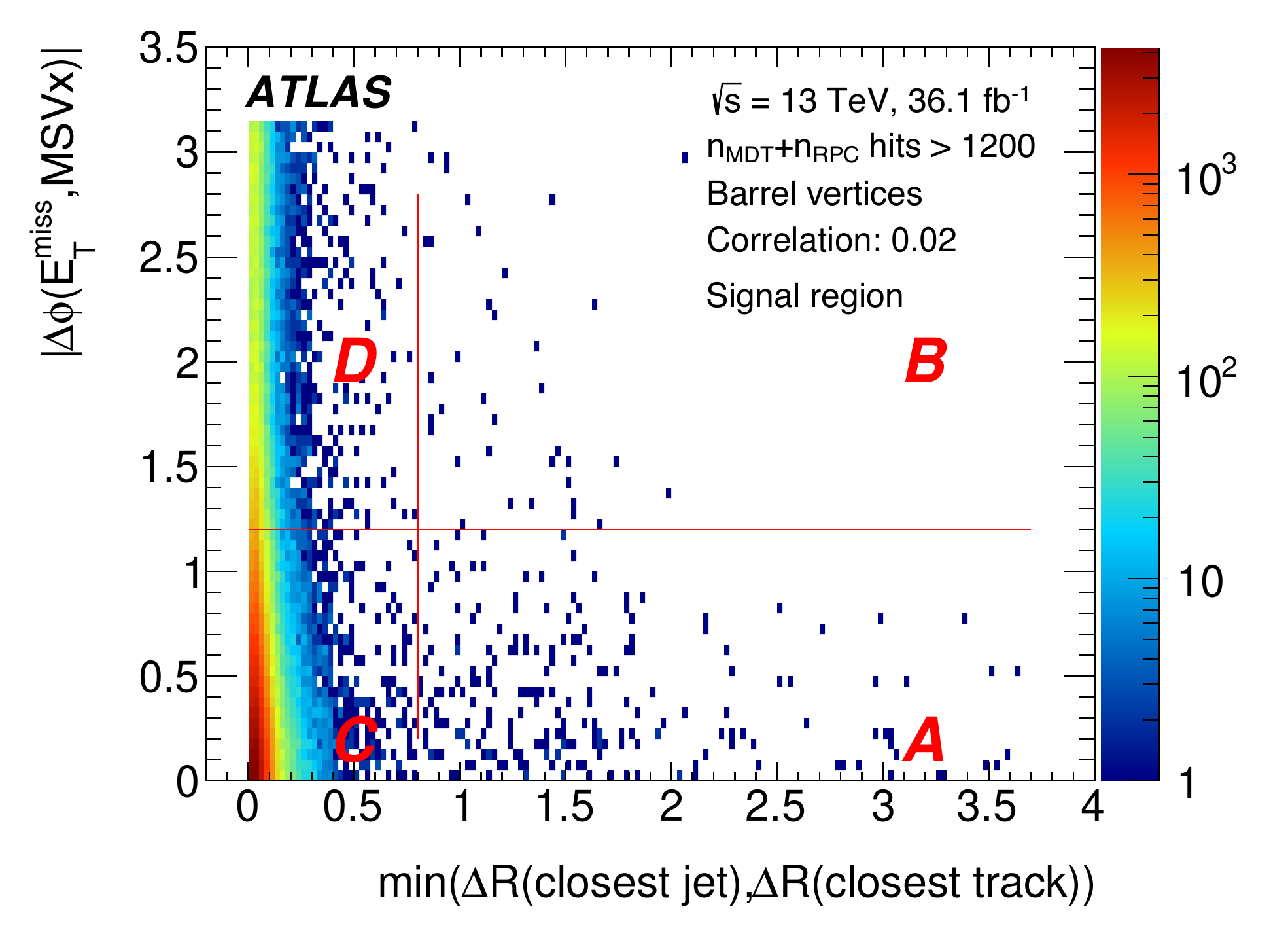}
	\caption{}
	\label{fig:ABCDplane_b_1MSVxMET}
	\end{subfigure}
	\caption{Distribution of barrel vertices in the ABCD plane for the SR and the definition of the four subregions for the (a) 1MSVx+Jets strategy and (b) 1MSVx+\MET strategy.}
	\label{fig:ABCDplane_b_1MSVx}
\end{figure}

The ABCD method developed for the 1MSVx+AO strategies uses two, uncorrelated vertex-based variables to create a two-dimensional plane that is split into four parts: region A is where most signal events are located, and three control regions (B, C, and D) that contain mostly background. The number of background events in A can be predicted from the population of the other three regions: $N_{\textrm{A}} = N_{\textrm{B}} \times N_{\textrm{C}}/N_{\textrm{D}}$, assuming negligible leakage of signal into regions B, C and D. This calculation is performed in two separate regions: one background-dominated validation region (VR) and one signal region (SR). Two different ABCD planes were defined for the 1MSVx+Jets and 1MSVx+\MET strategies. Figures~\ref{fig:ABCDplane_b_1MSVxJets} and \ref{fig:ABCDplane_b_1MSVxMET} show the distribution of barrel vertices for data in the ABCD plane and the definition of the four subregions for the 1MSVx+Jets and 1MSVx+\MET strategies, respectively.

The ABCD method relies on there being only one source of background, or multiple sources that have identical distributions in the ABCD plane.  In general, non-collision background, which does not originate from the $pp$ interaction point, will have a different distribution in the ABCD plane. To determine the non-collision background contamination, data collected in LHC empty bunch crossings throughout the 2016 data-taking period were used to estimate the number of non-collision background vertices in coincidence with events otherwise satisfying the single-vertex selection criteria. The empty bunch crossing trigger was not available in 2015, but the non-collision background's relative contribution is expected to be the same. The fraction of expected non-collision background vertices passing the final signal selection is negligible for the 1MSVx+Jets strategy while for the 1MSVx+\MET strategy it corresponds to 0.8\% (0.6\%) of the total number of background events expected in the SR in the barrel (endcaps). Non-collision background events are equally distributed in the ABCD plane and they are taken into account as a systematic uncertainty.

Signal contamination in the VR, which can bias the ABCD method validation, was tested and found to be negligible for both the1MSVx+Jets and 1MSVx+\MET strategies.

\subsubsection{ABCD plane for 1MSVx+Jets strategy}\label{sec:ABCDStealth}

For Stealth SUSY-like events the ABCD plane for background estimation is constructed with the isolation $\Delta R_{\text{min}}$ variable, and the sum of the numbers of MDT and trigger hits associated with the MS vertex, described in Section~\ref{sec:SigSelMSVxJets}. These two variables form the $x$-axis and $y$-axis of the ABCD plane, respectively.  The SR and the VR are built using the $\ET$ values of the leading and subleading jets and their definition is summarized in Table~\ref{tab:ABCDStealthCutflow}.

\begin{table}[!h]
\centering
\caption{Summary of the definition of the VR and SR used for the ABCD method for the 1MSVx+Jets strategy.}
\begin{tabular}{ll}
\toprule
Region & Criteria \\ 
\midrule
 \multirow{2}*{Barrel} & VR: $50 < E_{\textrm{T,subleading}} < 150$~\GeV, $E_{\textrm{T,leading}} > 150$~\GeV \\
                                  & SR: $E_{\textrm{T,leading}} > 150$~\GeV, $E_{\textrm{T,subleading}} > 150$~\GeV \\
\midrule 
\multirow{2}*{Endcaps} & VR: $100 < E_{\textrm{T,subleading}} < 250$~\GeV, $E_{\textrm{T,leading}} > 250$~\GeV \\
                                      & SR: $E_{\textrm{T,leading}} > 250$~\GeV, $E_{\textrm{T,subleading}} > 250$~\GeV \\
\bottomrule
\end{tabular}
\label{tab:ABCDStealthCutflow}
\end{table}
   
Signal contamination in regions B, C and D in the SR of the 1MSVx+Jets ABCD plane was found to be negligible. Signal contamination in the VR is negligible for benchmark samples with $m_{\tilde{g}} > 500$~GeV, and for $m_{\tilde{g}} = 500$~GeV in the barrel region. However, the endcaps region for $m_{\tilde{g}} = 500$~GeV and both barrel and endcaps regions for $m_{\tilde{g}} = 250$~GeV have non-negligible signal contamination in the VR and thus are not included in the 1MSVx+Jets strategy. Both the VR and SR show very low linear correlation between the two variables: $-$0.01 ($-$0.03) for the VR, and $-$0.01 ($-$0.05) for the SR in the barrel (endcaps). 

Table~\ref{tab:ABCDMethodStealth_VR} summarizes the observed and expected numbers of events in the four regions of the ABCD plane constructed using events from the VR. The number of observed events in region A is \mbox{$46$} and $11$ in the barrel and endcaps, respectively. These are in agreement with the \mbox{$45\pm5\textrm{ (stat.)}\pm9\textrm{ (syst.)}$} and \mbox{$15\pm3\textrm{ (stat.)}\pm12\textrm{ (syst.)}$} events predicted by the ABCD method in the barrel and endcaps, respectively. The systematic uncertainty associated with the background estimation reported above is described in detail in Section~\ref{ref:ABCDsyst}.

\begin{table}[htbp!]
\centering
\caption{Event counts in each of the four regions of the 1MSVx+Jets ABCD plane and expected number in region A obtained using 2015 and 2016 data from the VR. Both the statistical and systematic errors of the background expectation are reported.}\label{tab:ABCDMethodStealth_VR}
\begin{tabular}{lcccccc}
\toprule
VR  & A & Expected background & B & C & D  \\
\midrule
Barrel     & $46 $ & $45 \pm 5 \textrm{ (stat.)} \pm \,9 \textrm{ (syst.)}$ & $7,748 $ & $90 $ & 15,620  \\ 
Endcaps & $11 $ & $15 \pm 3 \textrm{ (stat.)} \pm 12 \textrm{ (syst.)}$ & $3,335 $ & $20 $ & $4,365 $ \\ 
\bottomrule
\end{tabular}
\end{table}

%
\subsubsection{ABCD plane for 1MSVx+\MET strategy}\label{sec:ABCDHiggs}

For the 1MSVx+\MET strategy the two variables used to define the ABCD plane are the isolation $\Delta R_{\text{min}}$ and the angle in the transverse plane between the \MET vector and the displaced vertex $|\Delta\phi(\textrm{\MET,MSVx})|$, described in Section~\ref{sec:SigSelMSVxMET}. The SR and VR are defined using the sum of the numbers of MDT and trigger hits in a cone around the MS vertex and their definition is summarized in Table~\ref{tab:ABCDHiggsCutflow}. Signal contamination in the VR is negligible.

Both the VR and SR show very low linear correlation between the two variables: 0.03 (0.01) for the VR, and 0.02 ($-$0.01) for the SR in the barrel (endcaps).

\begin{table}[!h]
\centering
\caption{Summary of the definition of the VR and SR used for the ABCD method for the 1MSVx+\MET strategy.}
\begin{tabular}{ll}
\toprule
Region & Criteria \\ 
\midrule
\multirow{2}{*}{Barrel}  & VR: $n_{\textrm{MDT}}$+$n_{\textrm{RPC}}<1200$ \\
                                     & SR: $n_{\textrm{MDT}}$+$n_{\textrm{RPC}}>1200$ \\

\midrule
\multirow{2}{*}{Endcaps}  & VR: $n_{\textrm{MDT}}$+$n_{\textrm{TGC}}<1500$ \\
                                         & SR: $n_{\textrm{MDT}}$+$n_{\textrm{TGC}}>1500$ \\
\bottomrule
\end{tabular}
\label{tab:ABCDHiggsCutflow}
\end{table}

Table~\ref{tab:ABCDMethodHSSData_VR} summarizes the observed and expected numbers of events in the four regions of the ABCD plane constructed using events from the VR. The number of observed events in region A is \mbox{$334\pm18$} and \mbox{$1107\pm33$} for the barrel and endcaps, respectively. These are in agreement with the \mbox{$319\pm29\textrm{ (stat.)}\pm38\textrm{ (syst.)}$} and \mbox{$1153\pm46\textrm{ (stat.)}\pm69\textrm{ (syst.)}$} events predicted by the ABCD method in the barrel and endcaps, respectively. The systematic uncertainty associated with the background estimation reported above is described in detail in Section~\ref{ref:ABCDsyst}.

\begin{table}[htbp!]
\centering
\caption{Event counts in each of the four regions of the 1MSVx+\MET ABCD plane and expected number in region A obtained using 2015 and 2016 data from the VR. Both the statistical and systematic errors of the background expectation are reported.}
\begin{tabular}{lcccccc}
\toprule
VR  & A & Expected background & B & C & D  \\
\midrule
Barrel      &    $334 $ &   $319 \pm 29$ (stat.) $\pm \,38$ (syst.)  & $119 $ & 67,980 & 25,380  \\
Endcaps  &  $1,107 $ & $1,153 \pm 46$ (stat.) $\pm \,69$ (syst.)  & $639 $ & 56,970 & 31,570  \\
\bottomrule
\end{tabular}
\label{tab:ABCDMethodHSSData_VR}
\end{table}

For the 1MSVx+\MET strategy the signal contamination in regions B, C and D of the SR ABCD plane is not negligible. The traditional method, which estimates the background in region A by taking the ratio of events in the adjacent regions, breaks down in the presence of signal in the control regions. An ABCD-likelihood method addresses this issue because it estimates the background in region A by fitting simultaneously background and expected signal events in the four regions of the ABCD plane. A likelihood function is formed from the product of four Poisson functions, one for each region A, B, C and D, describing signal and background expectations. The likelihood takes the form:
\begin{eqnarray}
& & \mathcal{L}(n_{\textrm{A}},n_{\textrm{B}},n_{\textrm{C}},n_{\textrm{D}} | s, b, \tau_{\textrm{B}}, \tau_{\textrm{C}}) = \prod_{i={\textrm{A,B,C,D}}} \frac{{\textrm{e}}^{-N_i} N_i^{n_i}}{n_i !} \quad , \nonumber
\end{eqnarray}
where $n_{\textrm{A}}$, $n_{\textrm{B}}$, $n_{\textrm{C}}$ and $n_{\textrm{D}}$ are the four observables that denote the number of events observed in each region in data. The $N_i$ are linear combinations of the signal and background expectation in each region, defined as follow:
\begin{eqnarray}
 & & N_{\textrm{A}} =  s + b \nonumber \\
 & & N_{\textrm{B}} = s \, \epsilon_{\textrm{B}} + b \, \tau_{\textrm{B}}  \nonumber \\
 & & N_{\textrm{C}} =  s \, \epsilon_{\textrm{C}} + b \, \tau_{\textrm{C}}  \nonumber \\
 & & N_{\textrm{D}} = s \, \epsilon_{\textrm{D}} + b \, \tau_{\textrm{B}} \, \tau_{\textrm{C}}   \nonumber 
\end{eqnarray}
where $s$ is the signal yield, $b$ the estimated background in region A, $\epsilon_i$ the signal contamination derived from MC simulation,  and $\tau_{\textrm{B}}$ and $\tau_{\textrm{C}}$ are the coefficients that relate the number of background events in region A to the other regions. The $s$, $b$ and $\tau_i$ values are allowed to float in the simultaneous fit to the four data regions.

\section{Systematic uncertainties} \label{sec:systematics}

In this section, experimental and theoretical systematic uncertainties associated with the signal predictions and background estimation are described.

\subsection{Uncertainties in the signal predictions}

The signal efficiency systematic uncertainties are dominated by the modeling of the signal physics processes, pileup and detector response and the extrapolation of the expected number of signal events as a function of the LLP proper lifetime.

One of the sources of systematic uncertainty associated with the Muon RoI Cluster trigger stems from the trigger scale factors, and it was estimated by moving them up and down by their statistical uncertainty. The trigger efficiency values obtained with these modified MC samples were compared with the efficiency of the nominal sample, and the difference was taken as the systematic uncertainty. A similar strategy was also adopted to estimate the trigger systematic uncertainty associated with the modeling of the minimum-bias interactions used to emulate pileup and the systematic uncertainty due to the PDF used to generate signal MC events. For the latter, the PDF uncertainty was obtained by considering the envelope of the uncertainty of the PDF set. The total systematic uncertainty of the signal efficiency for passing the Muon RoI Cluster trigger was obtained by summing in quadrature the contributions described above and varies from 1\% to 12\%, depending on the sample and detector region. The total systematic uncertainty of the signal efficiency for reconstructing an MS vertex was obtained by summing in quadrature the contributions coming from pileup and PDF uncertainties (evaluated with the similar procedure described above) and varies from 0.07\% to 5.5\%, depending on the sample and detector region. The systematic uncertainty associated with the prompt jets used for the 1MSVx+Jets strategy originates from jet energy scale, PDF and pileup uncertainties, and was evaluated with the similar procedure described earlier. The overall systematic uncertainty of the two-jet efficiency was determined by adding each component in quadrature and varies from 0.3\% to 9.8\%, depending on the sample and detector region. Systematic uncertainties due to the \MET computation for the 1MSVx+\MET strategy are negligible. By comparing the average number of muon segments in a cone around punch-through jets in data and MC simulation, systematic uncertainties associated with the mismodeling of the MS vertex reconstruction in signal events were found to be negligible.

For each of the scalar boson samples, excluding $m_{s} = 100$~\GeV, two proper lifetime points were fully simulated: one nominal sample and a secondary sample with longer proper lifetime, as described in Section~\ref{sec:samples}. The secondary sample is used to validate the extrapolation procedure, and a systematic uncertainty is assigned to each sample due to the non-closure of the extrapolation procedure. This is calculated by determining the fraction of events passing all analysis cuts in each MC sample generated with 9 m lab-frame decay length and comparing them to the expected global efficiencies obtained with the extrapolation procedure. The systematic uncertainty varies from 2\% to 37\%, depending on the sample. 
Stealth SUSY benchmark samples have events with kinematic behavior similar to that of events in the high-mass scalar boson samples and they are thus assigned the average systematic uncertainty of those samples (11\%). The kinematics of events in the baryogenesis samples is very close to the $\varPhi (125) \rightarrow ss$ kinematics and a comparable systematic uncertainty related to the extrapolation procedure is assumed. For that reason, for all the baryogenesis benchmark samples the average systematic uncertainty of the five $\varPhi (125) \rightarrow ss$ samples (32\%) is used.

The uncertainty in the combined 2015+2016 integrated luminosity is 2.1\%. It is derived, following a methodology similar to that detailed in Ref.~\cite{DAPR-2013-01}, using the LUCID-2 detector for the baseline luminosity measurements \cite{LUCID2}, from calibration of the luminosity scale using $x$--$y$ beam-separation scans.

\subsection{Uncertainties in the background prediction}\label{ref:ABCDsyst}

\begin{figure}[h!]
    \centering
    \includegraphics[width=0.48\textwidth]{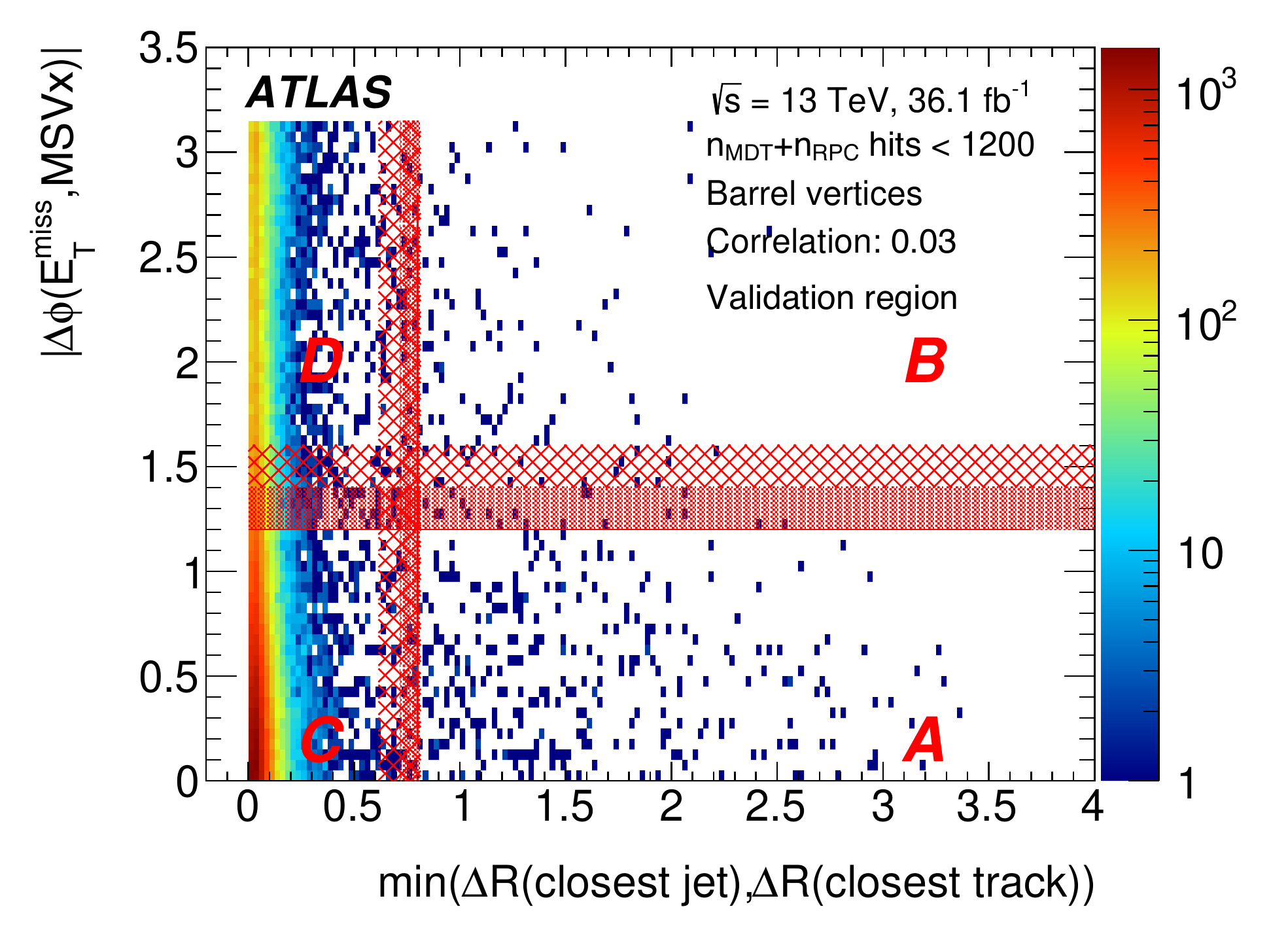}
    \caption{ABCD plane for the 1MSVx+\MET strategy in the barrel and definition of the regions excluded from the estimation of the background's systematic uncertainty.}
    \label{fig:ABCDsystSigmaBands}
\end{figure}

The systematic uncertainty associated with the background estimation of the ABCD method was evaluated using data events passing the validation region selection. Events falling in each of the two bands of the ABCD plane that surround region A, and that are shown in Figure~\ref{fig:ABCDsystSigmaBands} for the 1MSVx+\MET strategy in the barrel, were excluded and the expected background in the signal region was re-evaluated. The definition of region A was not modified in this procedure. The relative variation with respect to the observed events in region A was then evaluated. The size of the two bands is defined by the resolution of the two variables used to define the ABCD plane. The maximum value of the two results obtained by separately removing bands with widths of $1\sigma$ and $2\sigma$ was taken as the systematic uncertainty associated with the background estimation. Since the numbers of events in regions A and B are small, a bootstrap method~\cite{BootstrapMethod1,BootstrapMethod2} was used to determine the statistical uncertainty of the background estimate.

For the 1MSVx+Jets strategy the size of the two removed bands corresponds to 0.1 and 0.2 for the isolation and 100 and 200 for the number of MS hits. In the endcap validation region the uncertainty in the relative difference obtained from the bootstrap method is much higher than the statistical uncertainty of the background estimate in region A, while in the barrel the two are comparable. The final systematic uncertainty of the background estimate was taken to be the maximum of the statistical uncertainty from the bootstrap method and the statistical uncertainty from the background prediction in region A, and corresponds to 20\% in the barrel and 81\% in the endcaps.

For the 1MSVx+\MET strategy the size of the two removed bands corresponds to 0.1 and 0.2 for the isolation variable and 0.2 and 0.4 for $|\Delta\phi(\textrm{\MET,MSVx})|$. The relative difference in the validation region is lower than the statistical uncertainty on the relative difference obtained from the bootstrap method in both the barrel and endcaps. The final systematic uncertainty of the background estimate is taken to be the maximum of the statistical uncertainty from the bootstrap method and the statistical uncertainty of the background prediction in region A, corresponding to 12\% and 6\% for barrel and endcaps, respectively. The small contribution from non-collision background was taken into account as a systematic uncertainty of the background estimate, as discussed in Section~\ref{sec:ABCD}.

\section{Results}\label{sec:result}

For the 2MSVx strategy, $0.027\pm0.011$ background events are expected. After unblinding, no events passing the full signal selection were found.

For the 1MSVx+AO strategies, the number of observed events in the four regions of the ABCD plane and the background prediction in region A for events passing the SR selection are summarized in Table~\ref{tab:ABCDMethod_SR}. No significant excess above the predicted number of background events is found.

\begin{table}[htbp!]
\centering
\caption{Event counts in each of the four regions of the ABCD plane and expected number in region A for the SR, using the 2015 and 2016 datasets. Both the statistical and systematic errors of the background expectation are reported.}\label{tab:ABCDMethod_SR}
\begin{tabular}{clccccc}
\toprule
Strategy & Region & A & Expected background & B & C & D  \\
\midrule
\multirow{2}{*}{1MSVx+Jets} & Barrel     & 14 & $15 \pm 3$ (stat.) $\pm \,3$ (syst.)  & $2,057$ & $25$ & $3,414$ \\ 
                                               & Endcaps &   4 & $11 \pm 3$ (stat.) $\pm \,9$ (syst.)  & $560$ & $15$ & $761$ \\ 
\midrule
\multirow{2}{*}{1MSVx+\MET} & Barrel      &   224   & $243 \pm 38$ (stat.) $\pm \,29$ (syst.)  & $42 $  & 132,000  & 22,800  \\
                                                & Endcaps  &  489    & $497 \pm 51$ (stat.) $\pm \,30$ (syst.)  & $94$   & 165,800  & 31,390  \\
\bottomrule
\end{tabular}
\end{table}   

Upper limits on the production cross section times branching fraction were derived using the CL$_{\textrm{s}}$ prescription \cite{CLs}, implemented with the RooStat~\cite{cit:RooStat} and HistFactory~\cite{cit:HistFactory} packages using a profile likelihood function~\cite{Cowan:2010js}. For the 2MSVx and 1MSVx+Jets strategies the likelihood includes a Poisson probability term describing the total number of observed events. For the 1MSVx+\MET strategy the likelihood described in Section~\ref{sec:ABCDHiggs} was used. For scalar boson benchmark samples with $m_{\varPhi}\neq125$~\GeV, upper limits were set on $\sigma \times B$, where $B$ represents the branching fraction for $\varPhi \rightarrow ss$ assuming 100\% branching fraction into fermion pairs. For scalar boson benchmark samples with $m_{\varPhi}=125$~\GeV, upper limits were set on $\sigma/\sigma_{\textrm{SM}}\times B$, where $\sigma_{\textrm{SM}}$ is the SM Higgs boson production cross section, 48.58~pb \cite{HiggsXS13TeV}. For the Stealth SUSY benchmarks, upper limits were set on $\sigma/\sigma_{\textrm{SUSY}}\times B$, where $\sigma_{\textrm{SUSY}}$ is the SUSY production cross section for $pp \rightarrow \tilde{g}\tilde{g}$~\cite{SUSYXS13TeV} and $B$ represents the branching fraction for $\tilde{g}\rightarrow\tilde{S}g$, assuming that both $\tilde{S}\rightarrow S\tilde{G}$ and $S\rightarrow gg$ have 100\% branching fraction.

\begin{figure}[h!]
    \centering
    \begin{subfigure}[b]{0.51\textwidth}
        \includegraphics[width=\textwidth]{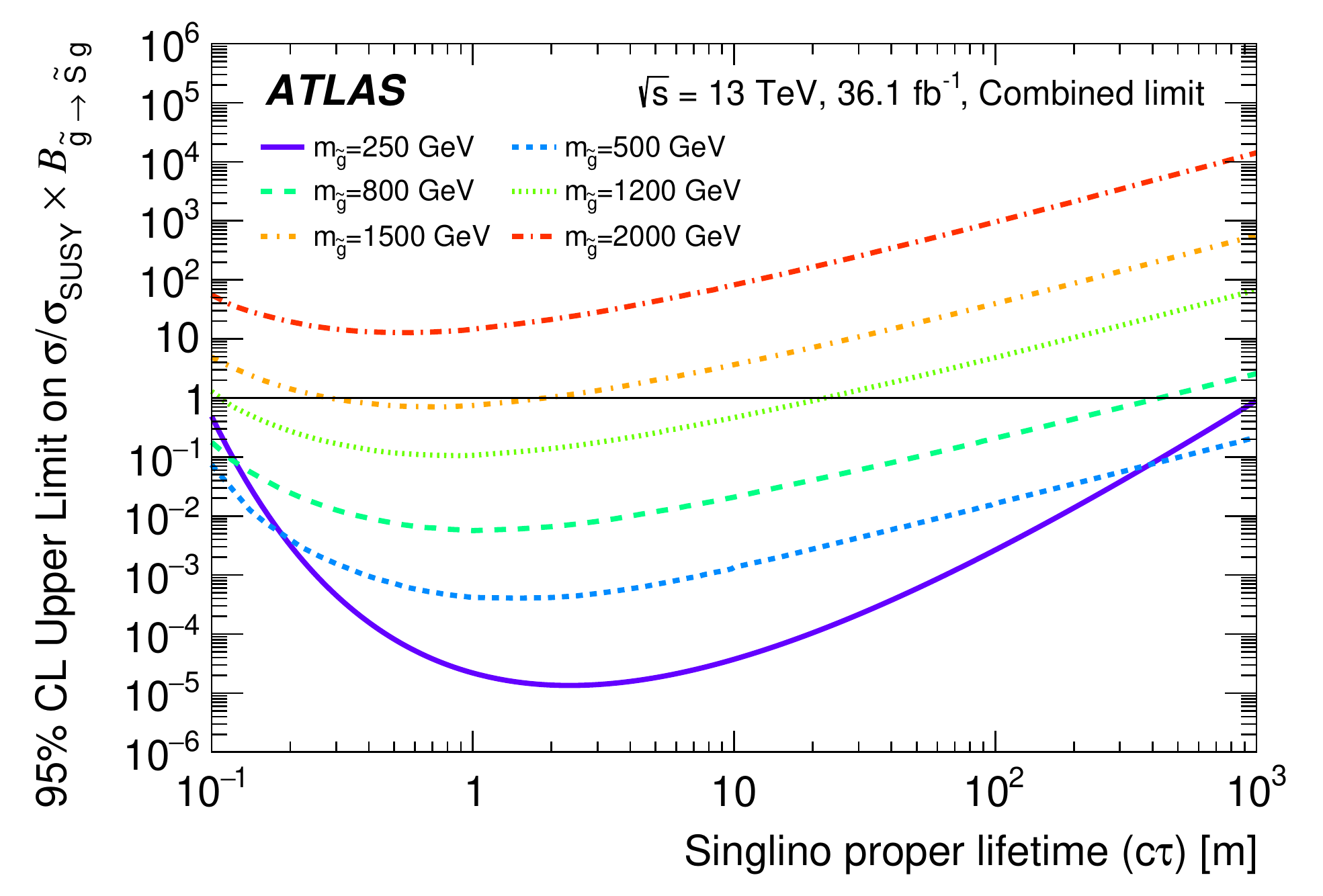}
        \caption{}
    \end{subfigure}
    \begin{subfigure}[b]{0.48\textwidth}
        \includegraphics[width=\textwidth]{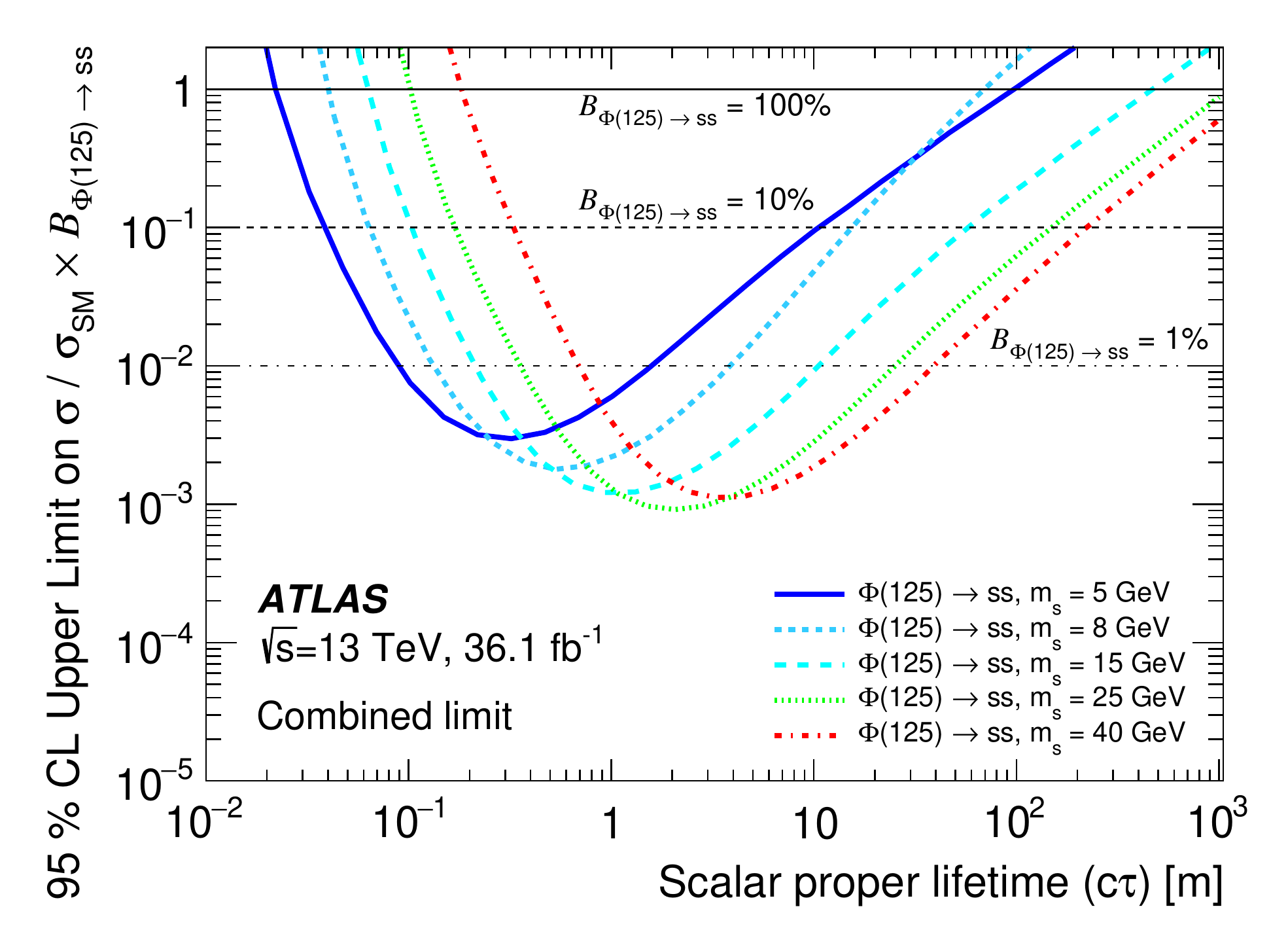}
        \caption{}
    \end{subfigure}
   \caption{Observed limits for (a) Stealth SUSY and (b) $\varPhi(125) \rightarrow ss$ benchmark samples obtained from the combination of 2MSVx and 1MSVx+AO strategies.}
    \label{fig:SummaryStealthHSS_CombExpLim}
\end{figure}

\begin{figure}
    \centering
        \includegraphics[width=0.48\textwidth]{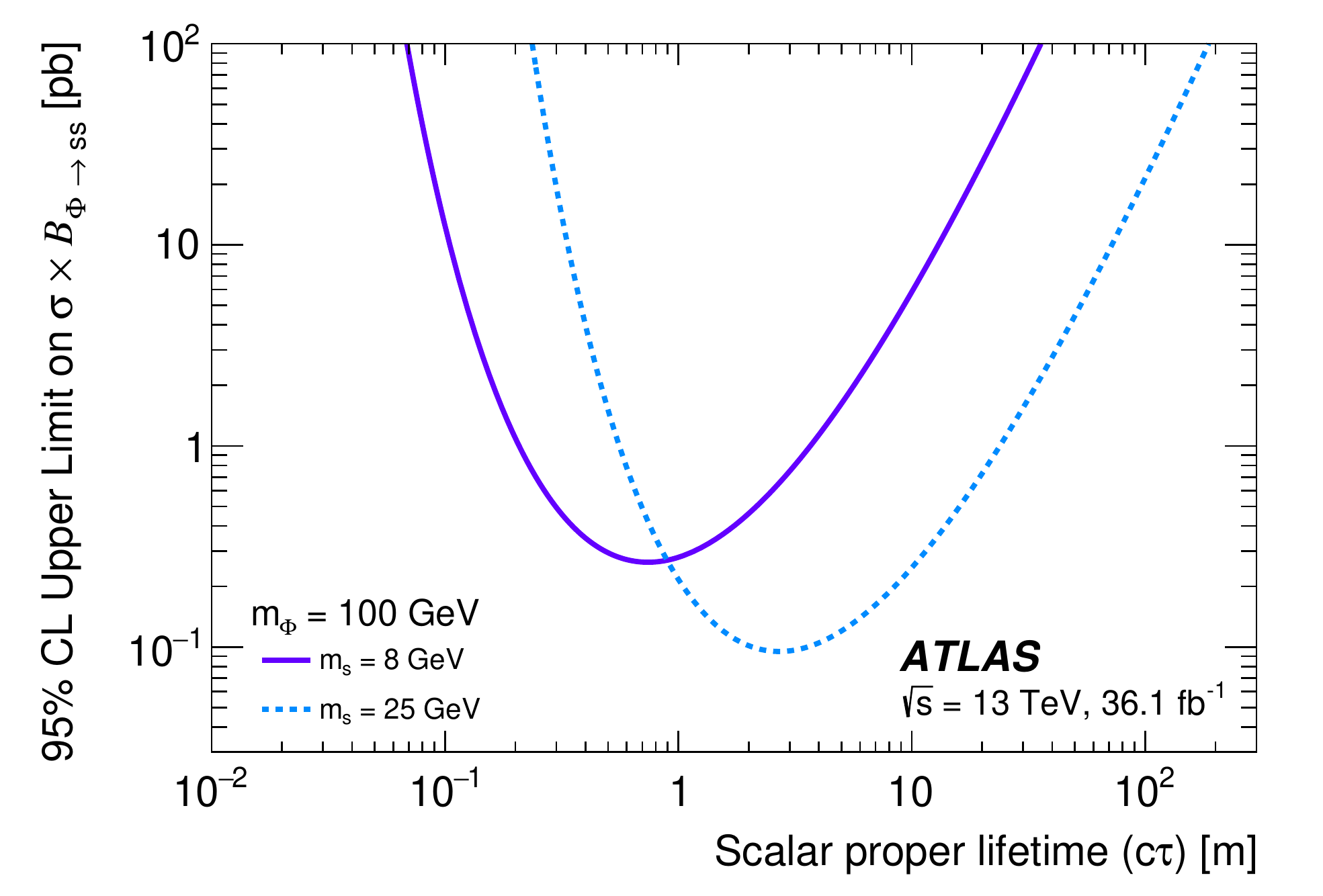}
        \includegraphics[width=0.48\textwidth]{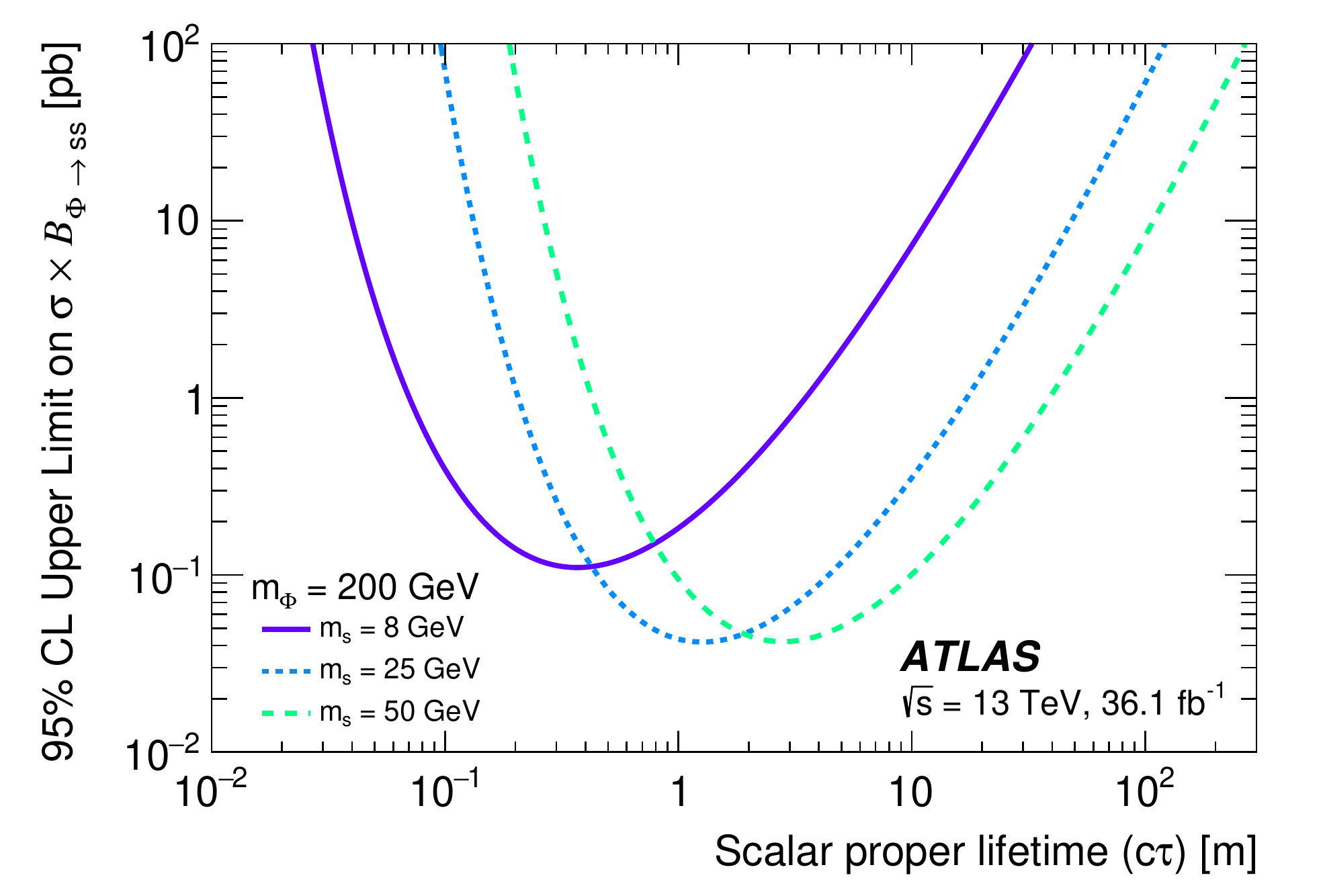}  \\
        \vspace{0.2 cm}
        \includegraphics[width=0.48\textwidth]{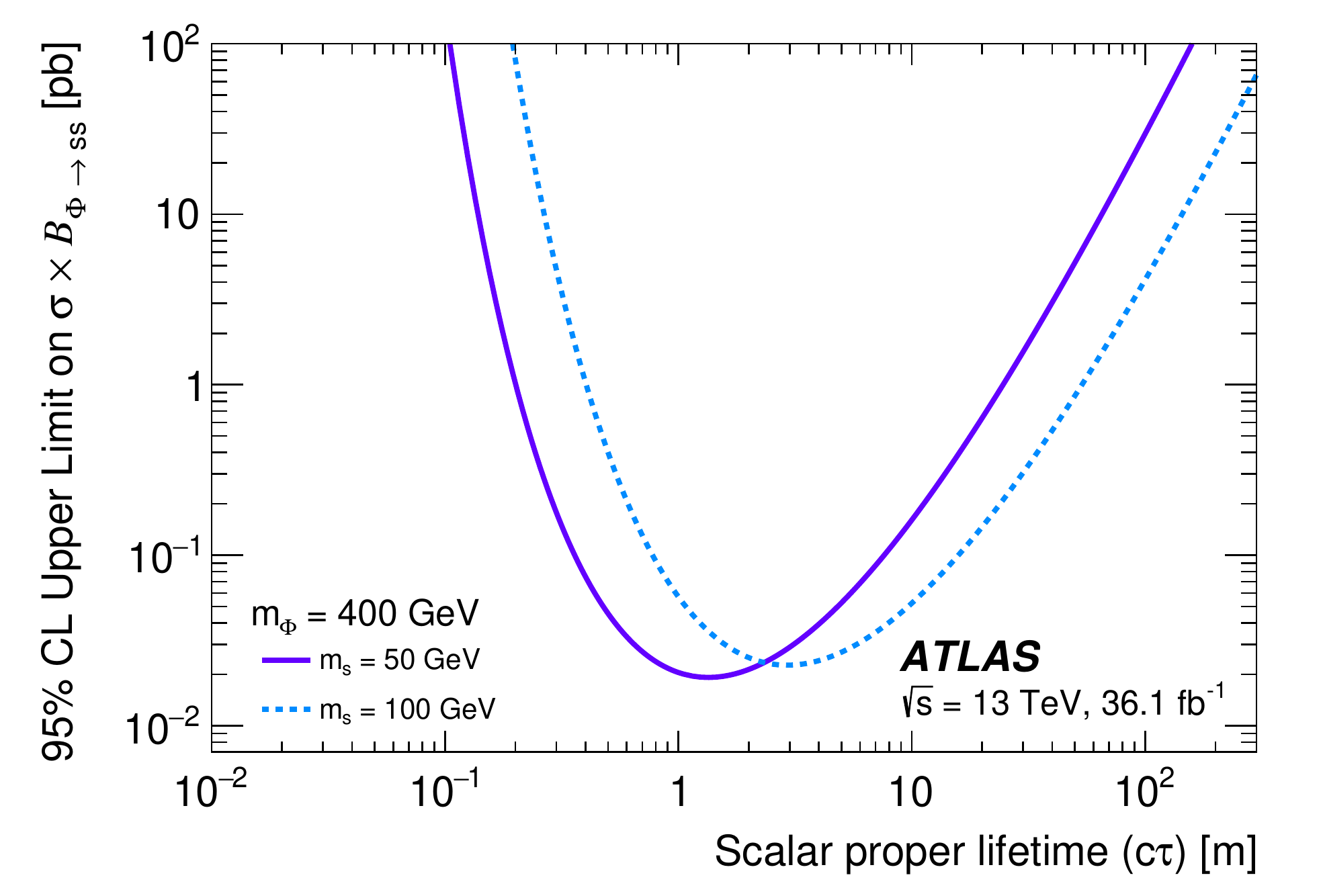}
        \includegraphics[width=0.48\textwidth]{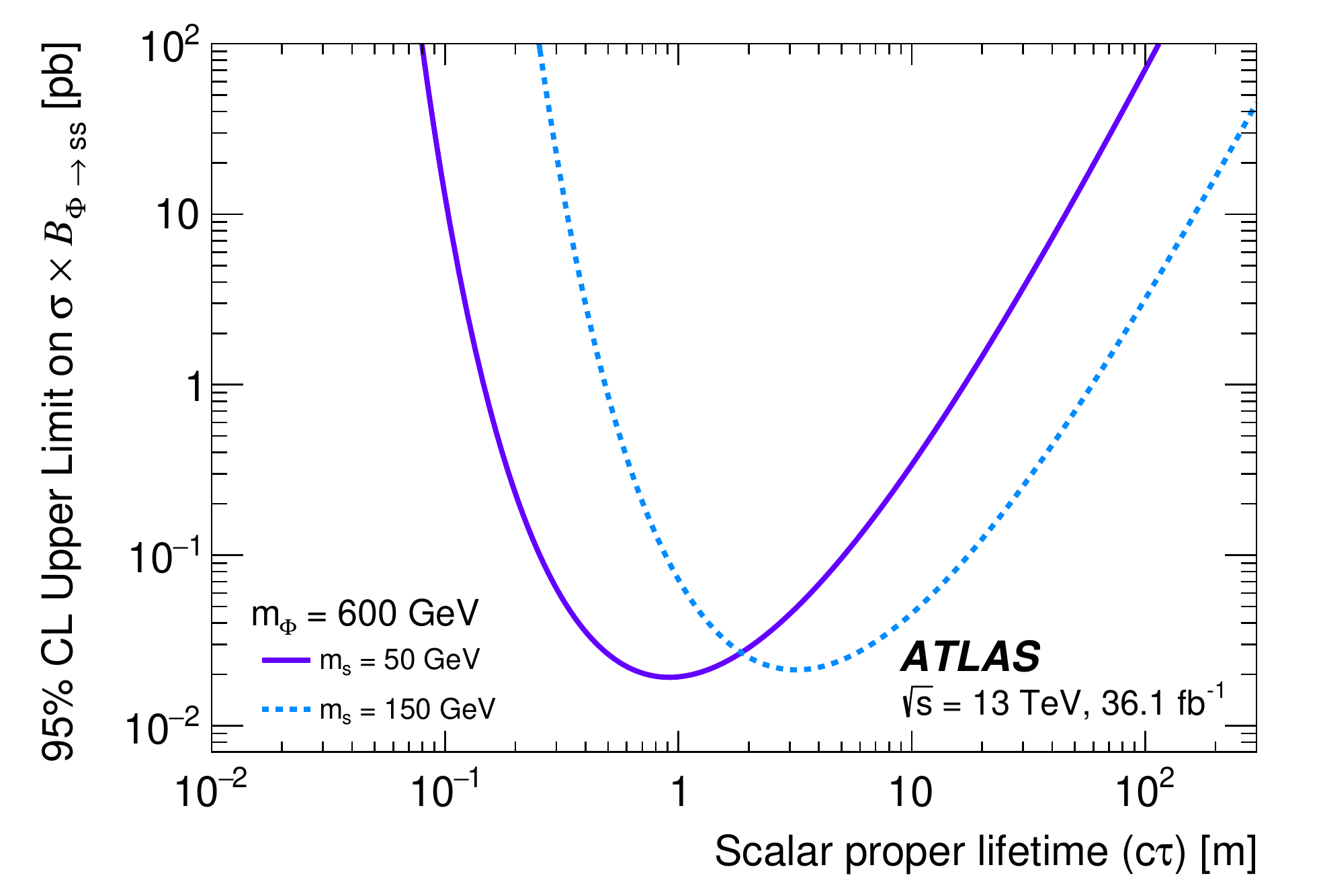} \\
        \vspace{0.2 cm}
        \includegraphics[width=0.48\textwidth]{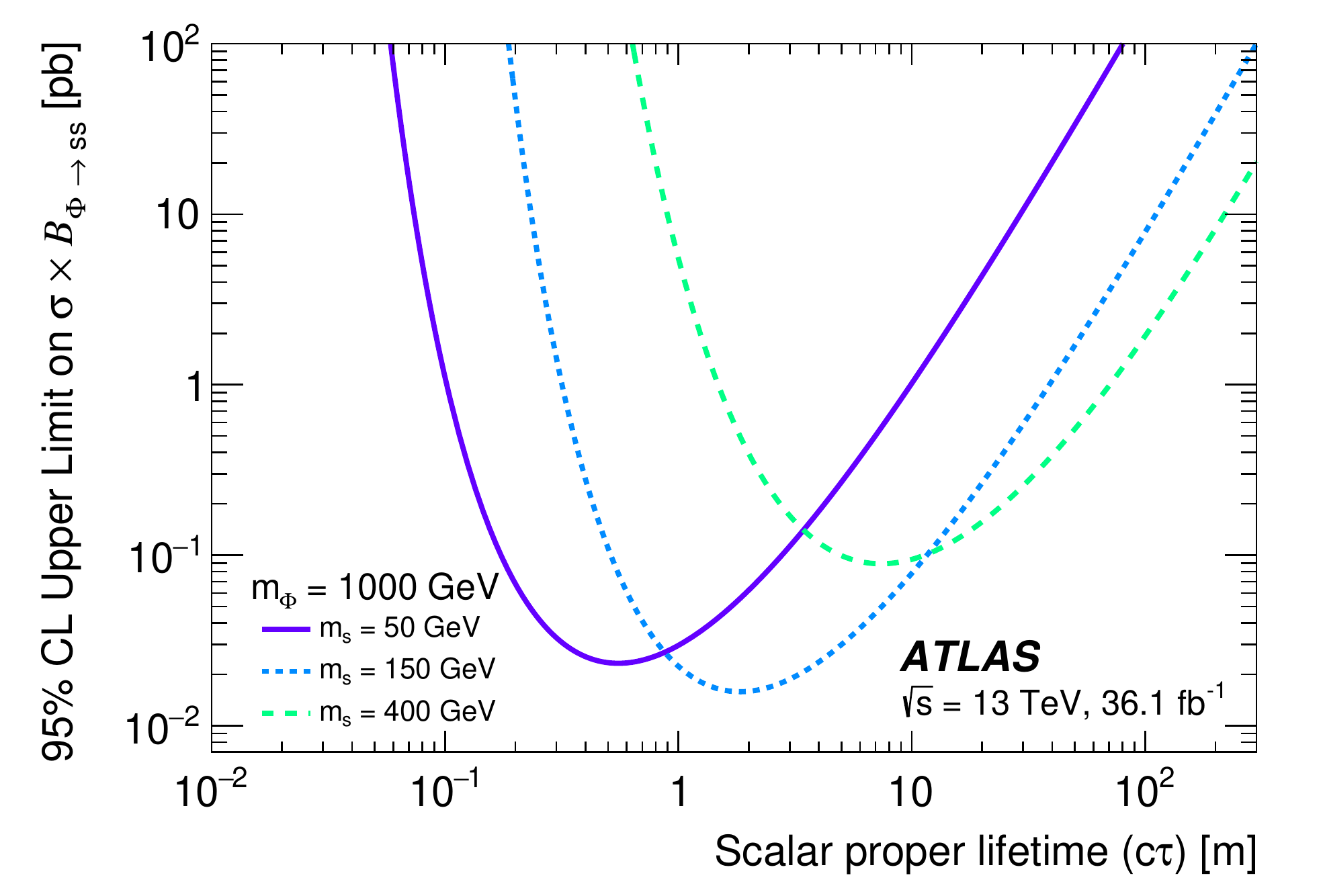}
   \caption{Observed limits for scalar boson benchmark samples with $m_\varPhi=100$~\GeV\ and $m_\varPhi>125$~\GeV. The different plots show the results obtained for different $\varPhi$ mass points.}
    \label{fig:Scalars}
\end{figure}

\begin{figure}[h!]
    \centering
    \begin{subfigure}[b]{0.48\textwidth}
        \includegraphics[width=\textwidth]{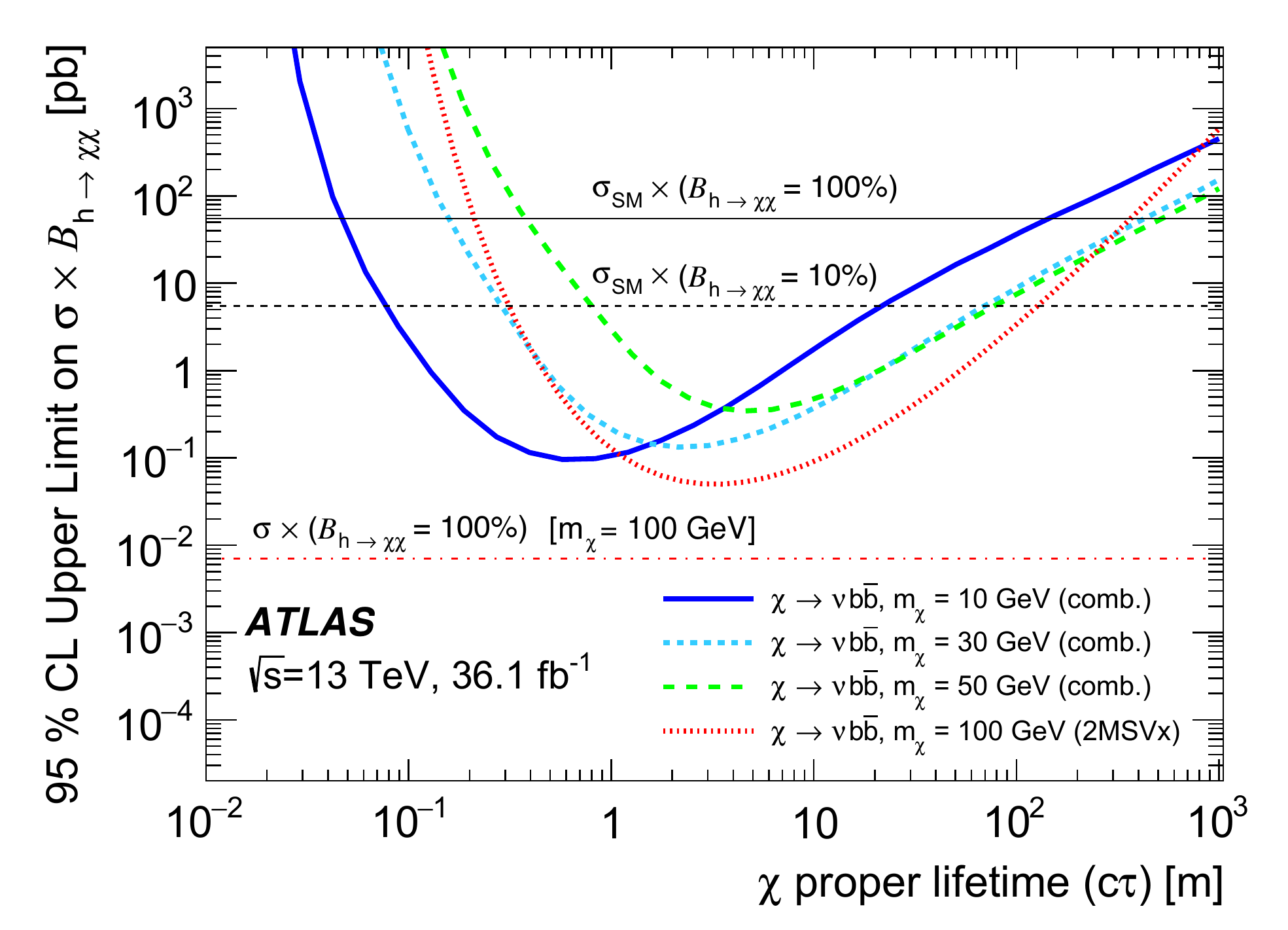}
        \caption{}
    \end{subfigure}
    \begin{subfigure}[b]{0.48\textwidth}
        \includegraphics[width=\textwidth]{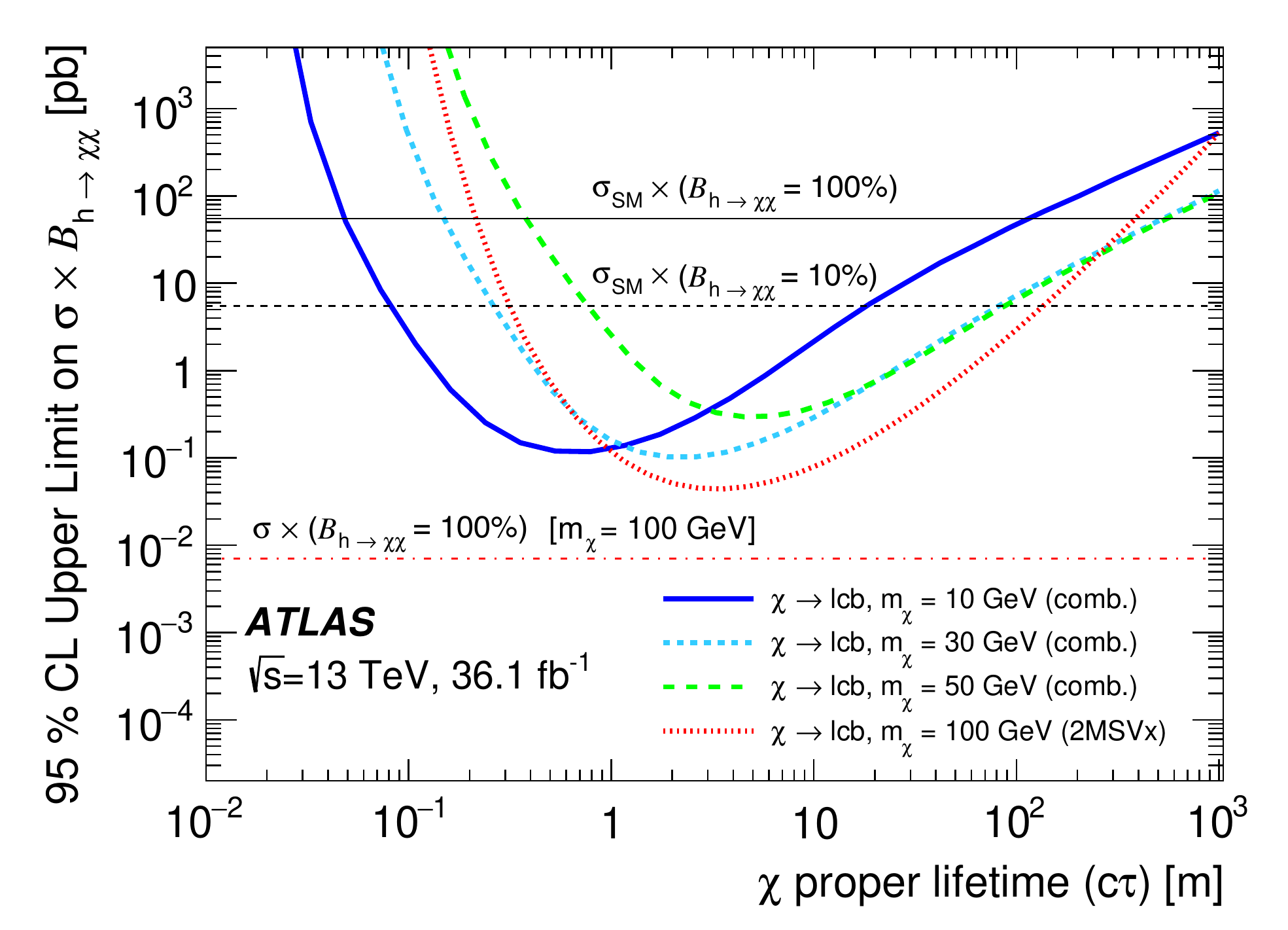}
        \caption{}
    \end{subfigure}
    \begin{subfigure}[b]{0.48\textwidth}
        \includegraphics[width=\textwidth]{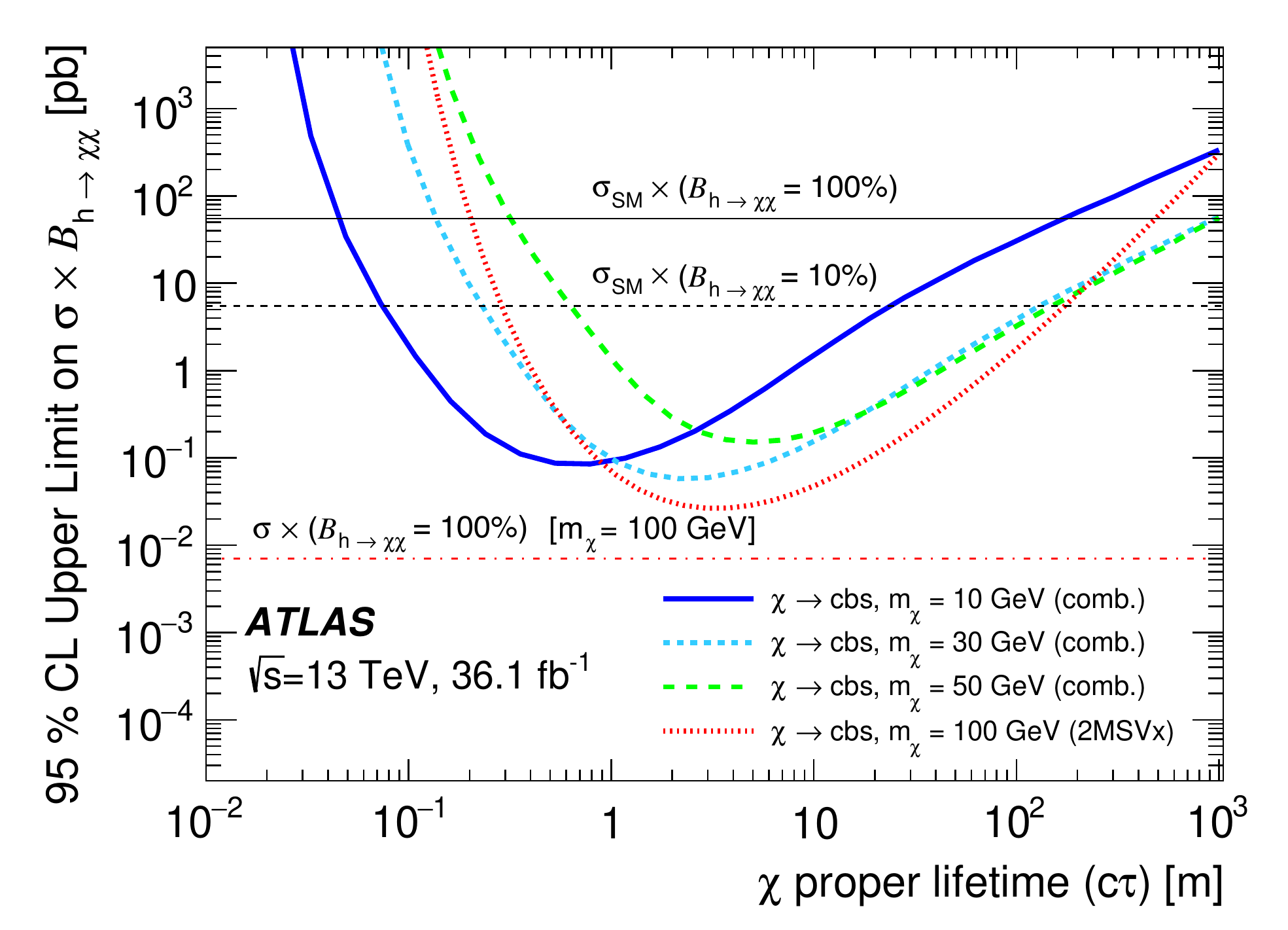}
        \caption{}
    \end{subfigure}
    \begin{subfigure}[b]{0.48\textwidth}
        \includegraphics[width=\textwidth]{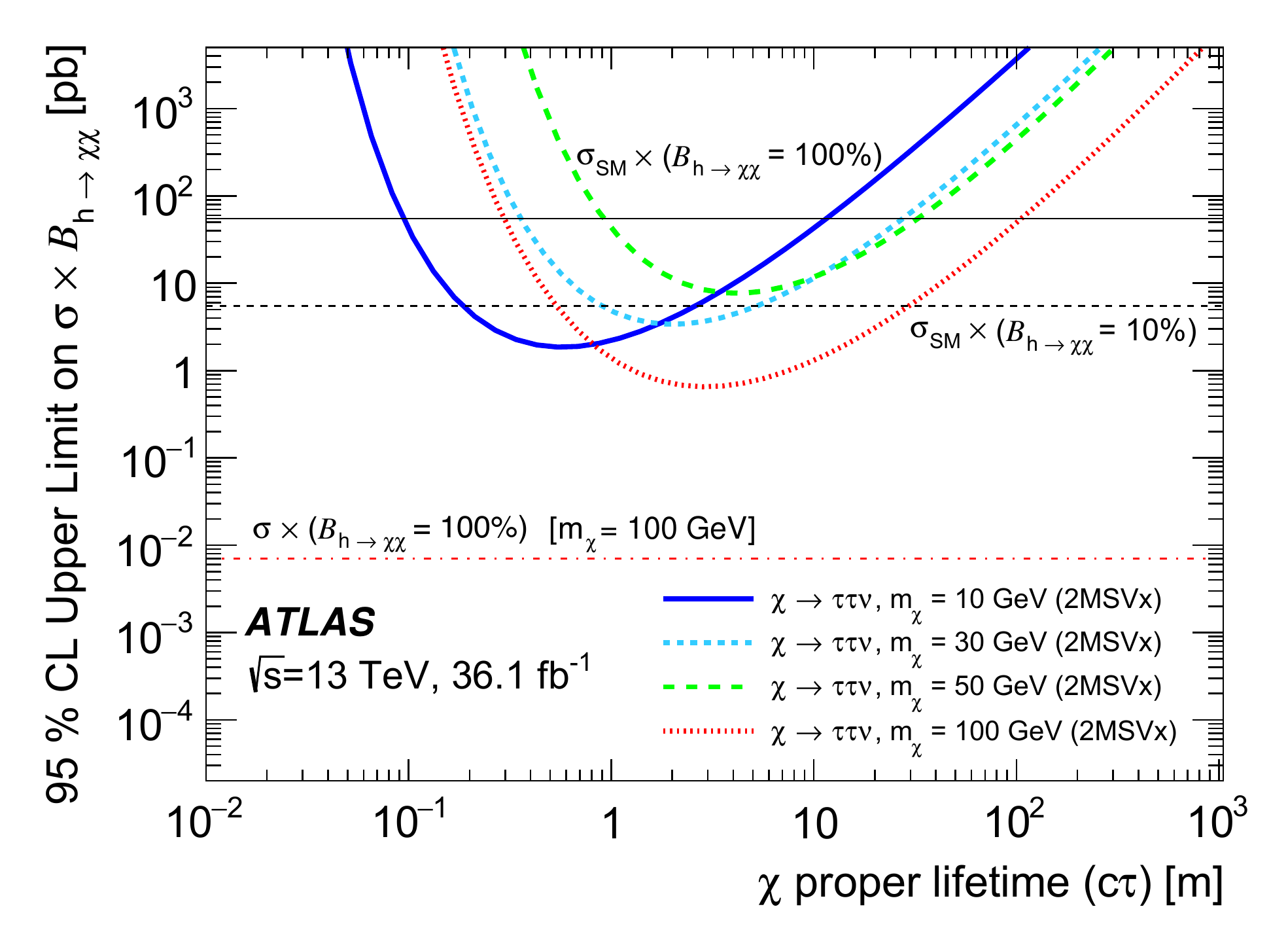}
        \caption{}
    \end{subfigure}
   \caption{Observed limits obtained for all the baryogenesis benchmark samples. Limits for the on-shell $h \rightarrow \chi\chi$ production in (a), (b) and (c) are obtained from the combination of 2MSVx and 1MSVx+\MET strategies, while for the off-shell sample ($m_\chi=100$~\GeV) the limits shown are obtained from the 2MSVx strategy only. Limits for the $\chi \rightarrow \tau\tau\nu$ channel shown in (d) are obtained from the 2MSVx strategy only. For reference, the black solid and dashed lines show respectively the 100\% and 10\% $\sigma \times B$ assuming the SM Higgs boson total production cross section, while the red dash-dotted line reports the 100\% $\sigma \times B$ for the off-shell $h \rightarrow \chi\chi$ production with $m_\chi=100$~\GeV, which is equal to 7~fb.}
    \label{fig:SummaryBaryo_CombExpLim}
\end{figure}

Figures~\ref{fig:SummaryStealthHSS_CombExpLim}, \ref{fig:Scalars} and \ref{fig:SummaryBaryo_CombExpLim} show the observed limits for all the MC benchmark samples considered in this paper. 
The limits were obtained from the combination of 2MSVx and 1MSVx+AO strategies, performing a simultaneous fit of the 2MSVx and 1MSVx+AO likelihood functions, except for the scalar boson samples with $m_{\varPhi}=100$~\GeV\ and $m_{\varPhi}>125$~\GeV, Stealth SUSY with $m_{\tilde{g}}=250$~\GeV~(both barrel and endcaps regions) and $m_{\tilde{g}}=500$~\GeV~(endcaps region), baryogenesis with $m_\chi=100$~\GeV\ and baryogenesis $\chi \rightarrow \tau\tau\nu$ benchmark samples. In these cases, the ABCD method developed for the analysis reported in this paper was found to be not optimal due to a large contamination by signal events in the VR or small signal--background separation for one of the variables of the ABCD plane. For those samples, the 2MSVx strategy provides strong limits and only those results are presented in this paper.

\begin{table}[htbp!]
\centering
\caption{Ranges of mean proper lifetime excluded at 95\% CL for scalar boson benchmark models with \mbox{$m_\varPhi=125$~\GeV}, assuming production cross-sections equal to 10\% or 1\% of the SM Higgs boson production cross-section~\cite{HiggsXS13TeV} for the combination of 2MSVx and 1MSVx+\MET strategies.}\label{tab:ExcludedRange_HSS}
\begin{tabular}{c|ccc}
\toprule
 $\varPhi(125) \rightarrow ss   $     &  \multicolumn{2}{c}{Excluded $c\tau$ range [m]} \\
$m_{s}$~[GeV]  &  10\%               & 1\% \\
\midrule
5   & 0.04--10.8  & 0.1--1.6 \\
8   & 0.07--15   & 0.14--3.8 \\
15 & 0.1--58     & 0.22--10.8 \\
25 & 0.2--149  & 0.4--25 \\
40 & 0.3--221  & 0.7--39 \\
\bottomrule
\end{tabular}
\end{table}

Table~\ref{tab:ExcludedRange_HSS} summarizes the lifetime ranges excluded by the analysis presented in this paper for branching fractions of 10\% and 1\% for the scalar boson with $m_\varPhi = 125$~\GeV\ decaying into two long-lived scalars. The results are substantially improved compared to the Run~1 analysis, where for 25~\GeV\ and 40~\GeV\ long-lived scalar masses the $c\tau$ ranges excluded for 1\% branching fraction were respectively 1.10--5.35~m and 2.82--7.45~m, while for lower long-lived scalar masses the Run~1 analysis did not have sensitivity at this level.

\clearpage

\section{Summary}\label{sec:conclusion}

This paper presents the results of a search for long-lived neutral particles using 36.1~fb$^{-1}$ of \mbox{proton--proton} collisions at $\sqrt{s}=13$~\TeV\ recorded at the LHC by the ATLAS detector in 2015 and 2016. Three separate strategies have been considered: two displaced vertices in the muon spectrometer (MS), one displaced vertex in the MS with two additional prompt jets, and one displaced vertex in the MS with $\MET>30$~\GeV. The observed number of events are consistent with the expected background and exclusion limits on the LLP production cross section as a function of its proper lifetime were computed for the theoretical benchmark models.

The results reported here are interpreted in terms of a scalar portal model similar to hidden-valley models where a boson can decay into two long-lived scalars, a Stealth SUSY model where each of the two long-lived singlino are produced from the gluino in association with a prompt jet, and a Higgs-boson-mediated baryogenesis model where the long-lived $\chi$ can decay into jets or leptons that violate the baryon and/or lepton number conservation. 

The two-vertex search was performed with the same strategy as adopted in Run\,1 and benefits from very low background, but at large $(c\tau)$ its sensitivity scales as $1/(c\tau)^2$, where $\tau$ is the proper lifetime of the LLP. The increased statistics and the analysis enhancements have improved the cross-section sensitivity for some of the $\varPhi \rightarrow ss$ decays by about an order of magnitude compared to the ATLAS Run\,1 analysis, and extended the sensitivity for the Stealth SUSY model to higher gluino masses that could not be reached with the Run\,1 search.

The one-vertex search extends the sensitivity, which scales as $1/(c\tau)$ at large $(c\tau)$, to much longer lifetimes. For the low-mass samples ($\varPhi(125) \rightarrow ss$ and Higgs portal baryogenesis) the sensitivity at shorter lifetimes is weaker than that attained with the two-MS-vertex search, while for the Stealth SUSY model in the high-mass regime ($m_{\tilde{g}}>500$~\GeV) the contribution of the one-MS-vertex search is dominant in the whole spectrum of proper lifetimes.

\section*{Acknowledgments}

We thank CERN for the very successful operation of the LHC, as well as the
support staff from our institutions without whom ATLAS could not be
operated efficiently.

We acknowledge the support of ANPCyT, Argentina; YerPhI, Armenia; ARC, Australia; BMWFW and FWF, Austria; ANAS, Azerbaijan; SSTC, Belarus; CNPq and FAPESP, Brazil; NSERC, NRC and CFI, Canada; CERN; CONICYT, Chile; CAS, MOST and NSFC, China; COLCIENCIAS, Colombia; MSMT CR, MPO CR and VSC CR, Czech Republic; DNRF and DNSRC, Denmark; IN2P3-CNRS, CEA-DRF/IRFU, France; SRNSFG, Georgia; BMBF, HGF, and MPG, Germany; GSRT, Greece; RGC, Hong Kong SAR, China; ISF and Benoziyo Center, Israel; INFN, Italy; MEXT and JSPS, Japan; CNRST, Morocco; NWO, Netherlands; RCN, Norway; MNiSW and NCN, Poland; FCT, Portugal; MNE/IFA, Romania; MES of Russia and NRC KI, Russian Federation; JINR; MESTD, Serbia; MSSR, Slovakia; ARRS and MIZ\v{S}, Slovenia; DST/NRF, South Africa; MINECO, Spain; SRC and Wallenberg Foundation, Sweden; SERI, SNSF and Cantons of Bern and Geneva, Switzerland; MOST, Taiwan; TAEK, Turkey; STFC, United Kingdom; DOE and NSF, United States of America. In addition, individual groups and members have received support from BCKDF, CANARIE, CRC and Compute Canada, Canada; COST, ERC, ERDF, Horizon 2020, and Marie Sk{\l}odowska-Curie Actions, European Union; Investissements d' Avenir Labex and Idex, ANR, France; DFG and AvH Foundation, Germany; Herakleitos, Thales and Aristeia programmes co-financed by EU-ESF and the Greek NSRF, Greece; BSF-NSF and GIF, Israel; CERCA Programme Generalitat de Catalunya, Spain; The Royal Society and Leverhulme Trust, United Kingdom. 

The crucial computing support from all WLCG partners is acknowledged gratefully, in particular from CERN, the ATLAS Tier-1 facilities at TRIUMF (Canada), NDGF (Denmark, Norway, Sweden), CC-IN2P3 (France), KIT/GridKA (Germany), INFN-CNAF (Italy), NL-T1 (Netherlands), PIC (Spain), ASGC (Taiwan), RAL (UK) and BNL (USA), the Tier-2 facilities worldwide and large non-WLCG resource providers. Major contributors of computing resources are listed in Ref.~\cite{ATL-GEN-PUB-2016-002}.

\printbibliography

\clearpage 
 
\begin{flushleft}
{\Large The ATLAS Collaboration}

\bigskip

M.~Aaboud$^\textrm{\scriptsize 34d}$,    
G.~Aad$^\textrm{\scriptsize 99}$,    
B.~Abbott$^\textrm{\scriptsize 125}$,    
O.~Abdinov$^\textrm{\scriptsize 13,*}$,    
B.~Abeloos$^\textrm{\scriptsize 129}$,    
D.K.~Abhayasinghe$^\textrm{\scriptsize 91}$,    
S.H.~Abidi$^\textrm{\scriptsize 164}$,    
O.S.~AbouZeid$^\textrm{\scriptsize 39}$,    
N.L.~Abraham$^\textrm{\scriptsize 153}$,    
H.~Abramowicz$^\textrm{\scriptsize 158}$
% ...
\bigskip
\\

% ...
$^{13}$Institute of Physics, Azerbaijan Academy of Sciences, Baku; Azerbaijan.\\
% ...
$^{34}$$^{(a)}$Facult\'e des Sciences Ain Chock, R\'eseau Universitaire de Physique des Hautes Energies - Universit\'e Hassan II, Casablanca;$^{(b)}$Centre National de l'Energie des Sciences Techniques Nucleaires (CNESTEN), Rabat;$^{(c)}$Facult\'e des Sciences Semlalia, Universit\'e Cadi Ayyad, LPHEA-Marrakech;$^{(d)}$Facult\'e des Sciences, Universit\'e Mohamed Premier and LPTPM, Oujda;$^{(e)}$Facult\'e des sciences, Universit\'e Mohammed V, Rabat; Morocco.\\
% ...
$^{39}$Niels Bohr Institute, University of Copenhagen, Copenhagen; Denmark.\\
% ...
$^{91}$Department of Physics, Royal Holloway University of London, Egham; United Kingdom.\\
% ...
$^{99}$CPPM, Aix-Marseille Universit\'e, CNRS/IN2P3, Marseille; France.\\
% ...
$^{125}$Homer L. Dodge Department of Physics and Astronomy, University of Oklahoma, Norman OK; United States of America.\\
% ...
$^{129}$LAL, Universit\'e Paris-Sud, CNRS/IN2P3, Universit\'e Paris-Saclay, Orsay; France.\\
% ...
$^{153}$Department of Physics and Astronomy, University of Sussex, Brighton; United Kingdom.\\
% ...
$^{158}$Raymond and Beverly Sackler School of Physics and Astronomy, Tel Aviv University, Tel Aviv; Israel.\\
% ...
$^{164}$Department of Physics, University of Toronto, Toronto ON; Canada.\\
% ...
$^{a}$ Also at Borough of Manhattan Community College, City University of New York, NY; United States of America.\\
$^{b}$ Also at California State University, East Bay; United States of America.\\
$^{c}$ Also at Centre for High Performance Computing, CSIR Campus, Rosebank, Cape Town; South Africa.\\
$^{d}$ Also at CERN, Geneva; Switzerland.\\
$^{e}$ Also at CPPM, Aix-Marseille Universit\'e, CNRS/IN2P3, Marseille; France.\\
% ...
$^{*}$ Deceased

\end{flushleft}


\end{document}